\documentclass[conference]{IEEEtran}
\IEEEoverridecommandlockouts
\AtBeginDocument{%
  }

\usepackage{graphicx}
\usepackage{cite}
\usepackage[hyphens]{url}
\usepackage{amsmath,amsfonts,amssymb}
\usepackage{algorithmic}
\usepackage{textcomp}
\usepackage{xcolor}
\usepackage{pifont}
\usepackage{circledtext}

\usepackage{graphicx}
\usepackage{booktabs}

\usepackage[most]{tcolorbox}

\definecolor{ACMPurple}{cmyk}{0.55,1,0,0.15}
\definecolor{ACMDarkBlue}{cmyk}{1,0.58,0,0.21}
\usepackage[
  bookmarks=true,
  breaklinks=true,
  colorlinks=true,
  linkcolor=ACMPurple,
  citecolor=ACMPurple,
  urlcolor=ACMDarkBlue,
  filecolor=ACMDarkBlue,
  pdfborder={0 0 0}
]{hyperref}

\usepackage{eso-pic}
\newcommand{\badgeurl}{https://doi.org/10.5281/zenodo.21542536}
\newlength{\badgewidth}
\newlength{\badgeskip}
\AddToShipoutPictureFG*{%
  \AtPageUpperLeft{%
    \raisebox{-\badgewidth-0.2in}{%
      \makebox[\paperwidth][r]{%
        \href{\badgeurl}{\includegraphics[width=\badgewidth]{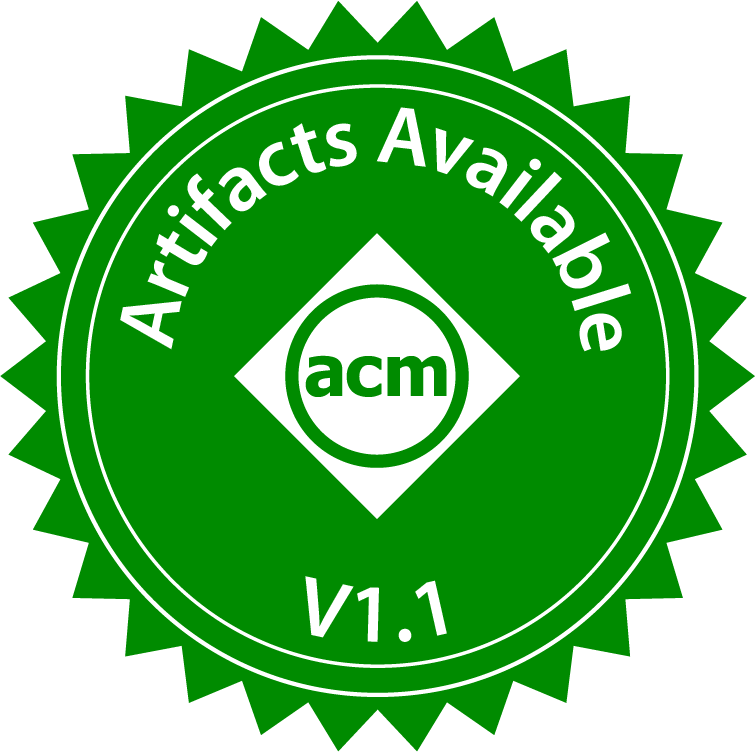}}\hspace{\badgeskip}%
        \href{\badgeurl}{\includegraphics[width=\badgewidth]{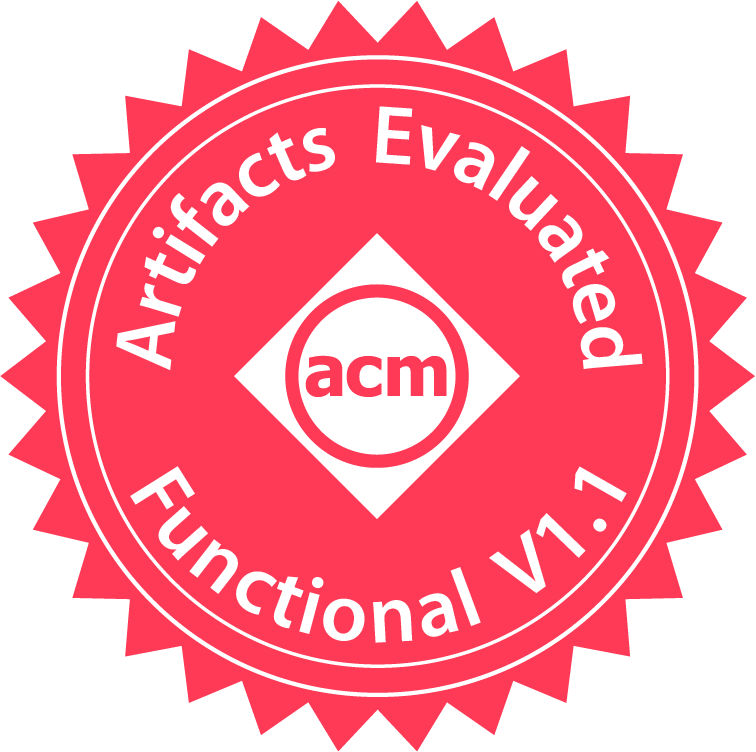}}\hspace{\badgeskip}%
        \href{\badgeurl}{\includegraphics[width=\badgewidth]{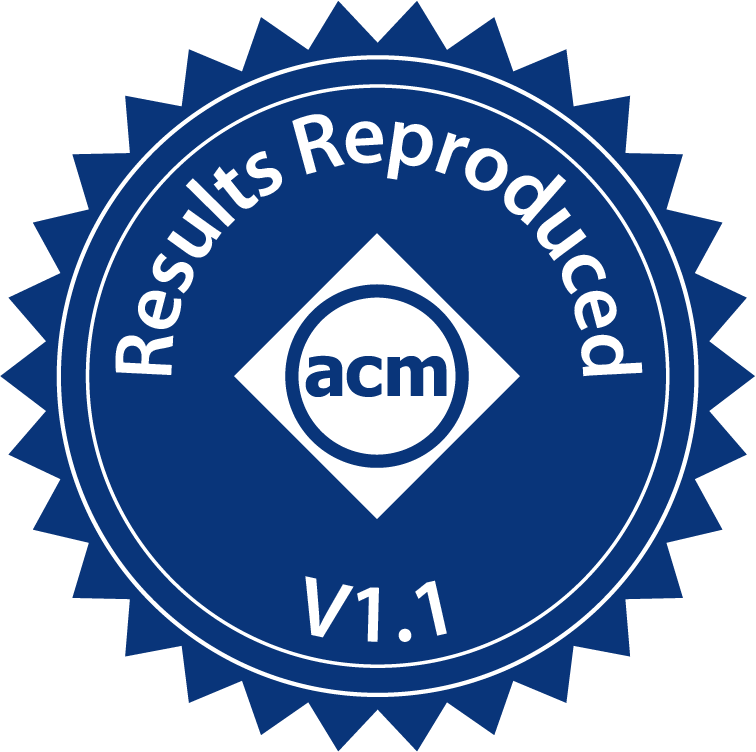}}%
        \hspace{0.75in}}}}}

\begin{document}

\title{Ray-by-Ray: Fine-Grained Resource Management for GPU Ray Tracing Units}

\author{\IEEEauthorblockN{Junrui Pan}
\IEEEauthorblockA{\textit{Purdue University} \\
West Lafayette, IN, USA \\
pan251@purdue.edu}
\and
\IEEEauthorblockN{Cesar Avalos Baddouh}
\IEEEauthorblockA{\textit{Purdue University} \\
West Lafayette, IN, USA \\
cavalosb@purdue.edu}
\and
\IEEEauthorblockN{Kai Ze Ee}
\IEEEauthorblockA{\textit{Purdue University} \\
West Lafayette, IN, USA \\
ee@purdue.edu}
\and
\IEEEauthorblockN{Timothy G. Rogers}
\IEEEauthorblockA{\textit{Purdue University} \\
West Lafayette, IN, USA \\
timrogers@purdue.edu}
}

%%%%%% -- PAPER CONTENT STARTS-- %%%%%%%%

\maketitle

\begin{abstract}
Ray tracing (RT) generates photorealistic images by simulating the physical behavior of light as it interacts with surfaces in a virtual scene. In film production and high-end rendering pipelines, this technique computes global illumination, reflections, refractions, and soft shadows by tracing millions of light rays per frame and accurately modeling their bounces and attenuation. Contemporary GPUs incorporate specialized ray tracing hardware to accelerate key operations in the traversal pipeline.
By accelerating these stages in hardware, the unit enables real-time ray tracing for complex scenes that previously required offline computation.
However, ray tracing introduces substantial control and data divergence. Each ray may traverse a different set of nodes in the acceleration structure, encounter distinct materials, or terminate at varying depths. 
This irregularity reduces thread coherence and exposes a fundamental disparity between the GPU SIMT and per-thread execution model of dedicated RT units, where each ray progresses independently.

In this work, we seek to validate assumptions made in open-source infrastructure and identify potential inefficiencies in resource management within the ray tracing pipeline. To address inefficiency caused by mismatches between SIMT and RT granularities, we propose Ray-by-Ray, which decomposes on-chip buffers into the granularity of individual rays and introduces lightweight management structures that enable more warps to run in parallel without increasing buffer capacity. This approach improves ray tracing unit utilization and increases memory-level parallelism. Experimental results show that Ray-by-Ray achieves an average performance improvement of 1.91x over the baseline, with an area overhead of 0.19\%.
\end{abstract}


\section{Introduction} \label{introduction}
% Ray tracing is capable of rendering photorealistic images by physically simulating how light interacts with objects in a scene, producing accurate lighting effects such as reflections, refractions, soft shadows, and global illumination. Unlike rasterization, which approximates lighting through precomputed or screen-space effects, ray tracing models the actual paths that light rays take as they travel, bounce, and scatter within a virtual environment. This physical accuracy has made ray tracing the core rendering technique in contemporary production pipelines {citations?}, where it is used to generate realistic images for visual effects and animation. Recent advances in GPU architecture have brought ray tracing into real-time applications by integrating dedicated ray tracing units that accelerate key operations in the traversal pipeline. By handling these operations in hardware, contemporary GPUs can now perform real-time ray tracing in complex, dynamic scenes that previously required hours of offline computation.

Ray tracing (RT) is capable of rendering photorealistic images by physically simulating how light interacts with objects in a scene, producing accurate lighting effects such as reflections, refractions, soft shadows, and global illumination. Unlike rasterization, which approximates lighting using precomputed or screen-space effects, ray tracing models the paths that light rays take as they travel, bounce, and scatter within a virtual environment. This physical accuracy has made ray tracing the foundation of production rendering pipelines~\cite{fascione_path_tracing_2018, droske_path_tracing_2023, georgiev_arnold_bruteforce_2018, christensen_ray_tracing_2006}, where it is widely used to generate realistic imagery for visual effects and animation. 

Recent advances in graphics processing unit (GPU) architecture have enabled real-time ray tracing by integrating dedicated hardware units that accelerate key operations in the traversal pipeline. By offloading these operations to specialized hardware, GPUs can now perform ray tracing in complex, dynamic scenes at interactive frame rates, tasks that previously required hours of offline computation. Even with hardware support for ray tracing operations, achieving real-time performance remains challenging. This is because ray tracing is highly divergent. 
%There are two main types of divergence in ray tracing: \textit{execution divergence} and \textit{data divergence}. 

% talk about bvh
% Right now industries trace rays by traversing a tree-like acceleration structure, Bounding Volume Hierarchy (BVH). BVH organizes geometries in the scene into a hierarchical tree. Because of the hierchical nature, missing a node guarantees missing in the child of the node, allowing large groups of objects to be quickly excluded when a ray misses in a node of the tree. To speed up BVH traversal, GPU vendors introduced dedicated ray tracing units (RT units) to accelerate long running operations in the traversal pipeline, such as ray-box intersection tests, ray-primitive intersections, and instance transformations. 

% then issues with RT
\begin{figure}[h]
    \centering
    \includegraphics[width=\linewidth]{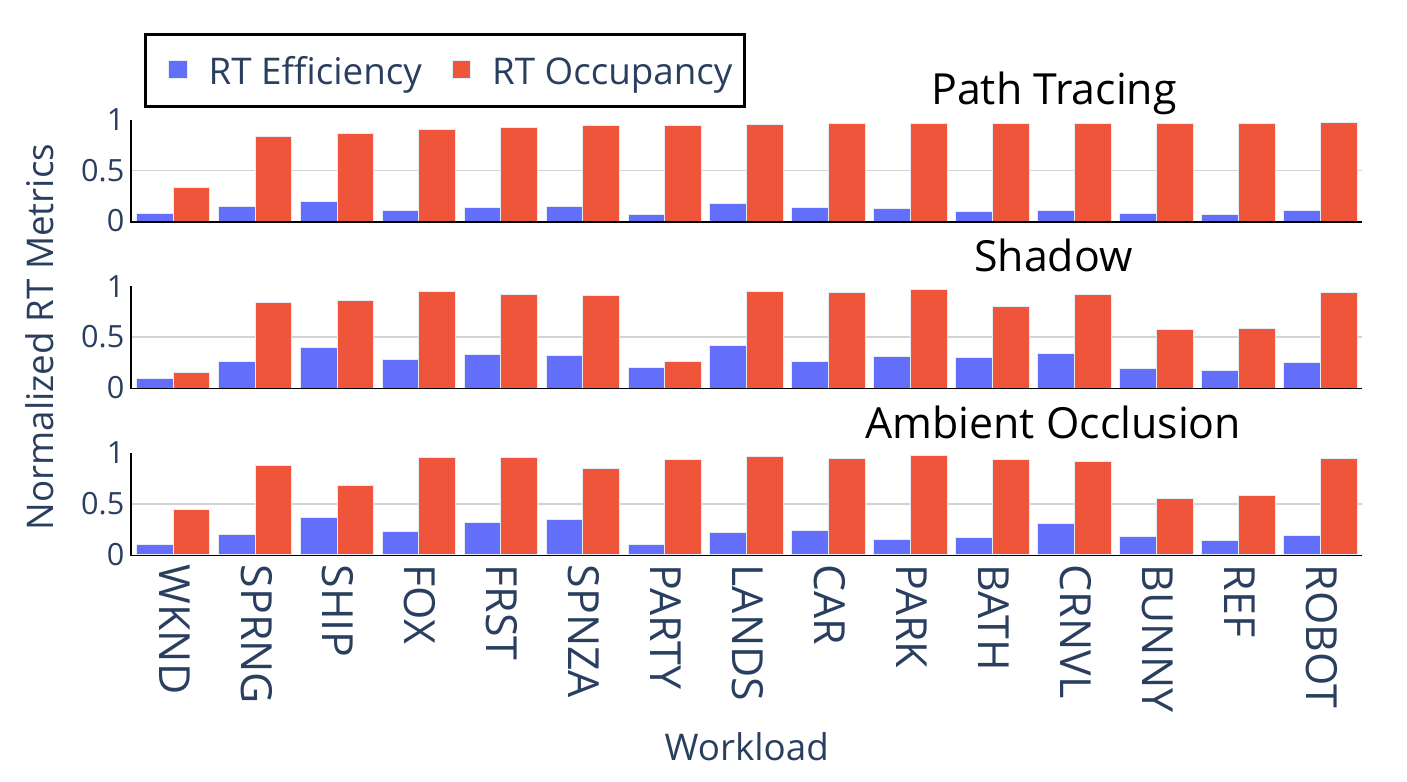}
    % or figure* + \pagewidth for bigger images
    \caption{Simulated RT occupancy and efficiency using Vulkan-Sim~\cite{saed_vulkansim_2022, liu_lumibench_2023}. Occupancy is the average ratio of warps resident in the RT unit to its maximum allowed warps; efficiency is the average ratio of threads active in the RT unit to its maximum thread capacity.}
    \label{fig:low_eff_high_occ}
\end{figure}

Contemporary GPUs~\cite{amd_rdna4_instruction_2025,barczak_intel_arc_,nvidia_nvidia_ampere_} perform ray traversal by using a tree-like acceleration structure called a Bounding Volume Hierarchy (BVH). The BVH organizes scene geometry into a hierarchical tree, where each node represents a bounding volume that encloses its child nodes. If a ray misses a parent node, it misses all children, efficiently excluding large object groups from further intersection tests. To perform the traversal of these BVH trees in hardware, patents from NVIDIA~\cite{muthler_techniques_traversing_2022} and state-of-the-art academic work~\cite{saed_vulkansim_2022, ha_generalizing_ray_2024,tozlu_cooprt_2025} assume a single CISC instruction is issued from each warp to perform the traversal of all the 32 rays within the warp. GPUs follow a Single-Instruction, Multiple-Thread (SIMT) execution model, where a warp of 32 threads executes in lockstep. However, this model does not map well to BVH traversal~\cite{aila_understanding_2009}, motivating the introduction of dedicated RT cores. All threads within a warp enter the RT core together, perform the traversal, and return as a group once the traversal has fully completed.
Each ray traverses the scene independently and may intersect different objects. Some rays descend deeper into the BVH while others terminate early. This mismatch between warp-level management and thread-level behavior, where each ray progresses differently, creates \textit{control-flow divergence}.
As a result, rays that complete early are forced to wait for the longest-running rays within the same warp. 
% Hardware RT cores exhibit the same behavior: although each thread performs traversal independently, the warp cannot complete the RT instruction until all rays finish.

% Divergence also continues beyond traversal. After traversal completes, each ray may invoke different shader programs (e.g., closest-hit, miss, or any-hit shaders) based on its intersection results, introducing further execution imbalance across threads.

% Even when all threads within a warp invoke the same hit shader and avoid execution divergence, \textit{data divergence} can still occur. Each ray may intersect a different object and fetch distinct textures or material data, leading to incoherent memory accesses and reduced cache efficiency. 
Fig.~\ref{fig:low_eff_high_occ} shows that although overall RT unit occupancy is high, indicating that sufficient warps are available to keep the RT units busy, RT efficiency remains low. RT efficiency is defined as the fraction of threads actively performing traversal work. This low efficiency arises because rays follow different paths during traversal, and some terminate much earlier than others. The problem is particularly severe in path tracing workloads: when a ray hits an object, a new ray is launched from the hit point to simulate reflection. This process repeats until the ray terminates and invokes shading. The recursive nature of path tracing amplifies execution divergence and further degrades overall ray tracing efficiency. 

To help reduce the effects of threads within a warp traversing different paths, NVIDIA introduced Shader Execution Reordering (SER) \cite{nvidia_shader_execution_}. SER allows the GPU to dynamically regroup threads based on their execution state into new, coherent warps. However, this mechanism requires writing out the full execution context of each thread and later restoring it when the thread is reassigned to a different warp. This context save-and-restore operation introduces overhead, and the extra memory traffic and latency limit its effectiveness~\cite{nvidia_shader_execution_,nvidia_improve_shader_2022,nvidia_path_tracing_2025}. Because SER operates strictly as a post-traversal mechanism, it is incapable of resolving divergence during the active ray traversal phase.
% As a result, SER must be applied selectively and is primarily effective in highly divergent workloads, such as path tracing~\cite{}. 

% Hardware-based techniques that improve SIMT efficiency~\cite{fung_dynamic_2007,fung_thread_2011} avoid the context switching overhead, but they rely on explicit divergence tracking. In contrast, ray traversal has clear divergence boundaries—threads diverge when entering the RT core and reconverge only when exiting it—so the instruction-level divergence these mechanisms target does not arise in hardware-accelerated traversal.

% While numerous academic works have analyzed and proposed improvements to dedicated ray tracing cores in state-of-the-art GPUs \cite{chou_treelet_prefetching_2023,chou_treelet_accelerated_2025,tozlu_cooprt_2025,liu_intersection_2021,jung_hierarchical_2025}, none have examined how existing hardware is actually designed. In this work, we perform a detailed study and microbenchmarking analysis of contemporary NVIDIA RT cores. Our findings reveal several key inefficiencies in the current RT core design that arise from suboptimal resource management. Specifically, on-chip resources within the RT cores are allocated at warp granularity, creating waste when lanes within a warp are inactive, which is commonly observed in RT workloads.
While numerous academic works have analyzed and proposed improvements to dedicated ray tracing cores in state-of-the-art GPUs \cite{chou_treelet_prefetching_2023,chou_treelet_accelerated_2025,tozlu_cooprt_2025,liu_intersection_prediction_2021,jung_hierarchical_2025}, none of these studies recognized the underlying inefficiency caused by warp-granular resource allocation. 
% In this work, we examine this behavior directly on contemporary NVIDIA RT cores, and our analysis identifies the root cause of the inefficiency. 
The assumption of all these prior works is that RT resources are allocated and deallocated at a warp granularity. To validate this assumption, we perform microbenchmarking on the RT core using the OptiX interface and a cycle-level analysis tool we developed. From this analysis, we observe that on-chip resources inside contemporary RT cores are allocated at warp granularity, causing substantial waste whenever lanes in a warp are inactive, which is a common scenario in ray tracing workloads. 
Unlike traditional SIMT code, ray traversal exhibits memory access patterns that gain little from coalescing; retaining warp-granular execution offers limited benefit. Moreover, our measurements show that the RT core operates as a relatively narrow pipeline, making it naturally amenable to sharing among threads from different warps rather than servicing an entire warp at once.
{
Building on the insights from our study, we propose decomposing each ray buffer group into finer-grained ray buffers paired with lightweight hardware for dynamic warp-to-thread mapping. Our guiding principle is to exploit divergence rather than eliminate it. To achieve this, we decouple software-visible SIMT divergence from hardware-level execution divergence. At the software level, we leave the standard control-flow divergence untouched: the programming model still sees the original active-lane mask carried by a warp, and a ray strictly returns to its original lane and warp once traversal completes. Instead, our design targets the internal hardware divergence that occurs inside the RT unit. This under-utilization arises because the unit traditionally manages buffers and traversal stacks at a warp granularity, even though ray traversal is inherently a thread-level operation. By dynamically reordering active rays onto available RT-unit resources, our hardware turns this internal divergence into reclaimable capacity rather than wasted occupancy. We can isolate this optimization because the entire traversal process is encapsulated within a single complex instruction, which bounds the divergence phase and creates a clean architectural boundary. Consequently, we can manage divergent execution entirely in hardware without the extensive overheads of general-purpose dynamic warp formation, which recomposes threads into entirely new warps~\cite{fung_dynamic_2007,fung_thread_block_2011a}. Ultimately, this allows more rays to traverse concurrently. Because ray traversal is heavily memory-latency bound, this increased memory-level parallelism successfully hides memory latency.
% With the insights from our study, we propose decomposing each ray buffer group into finer-grained ray buffers, paired with lightweight hardware that dynamically manages warp-to-thread mappings. A key principle is that we do not eliminate divergence; we exploit it. Software (SIMT) divergence is the control-flow divergence visible to the programming model: the active-lane mask a warp carries through the shader. We leave this untouched: a ray returns to its original lane and warp once traversal completes. What we target instead is hardware execution divergence inside the RT unit, the under-utilization that arises because the unit manages buffers and traversal stacks at warp granularity while traversal is inherently a thread-level operation. By dynamically reordering active rays onto RT-unit resources, our design turns this divergence into reclaimable capacity rather than wasted occupancy.
% This isolation is possible because the entire traversal is encapsulated in a single complex instruction, which bounds the divergence phase and provides a clean architectural boundary. We can therefore manage divergent execution in hardware without the extensive overheads of general-purpose dynamic warp formation, which recomposes threads into entirely new warps~\cite{fung_dynamic_2007,fung_thread_block_2011a}. The result is more rays traversing concurrently; since traversal is memory-latency bound, this added memory-level parallelism better hides memory latency.
}

In summary, we make the following contributions:
\begin{itemize}
    \item We propose Ray-by-Ray, a lightweight, fine-grained per-ray resource management mechanism within the ray tracing unit to better utilize on-chip resources, enabling higher ray parallelism to hide memory latency. Across a wide range of workloads, this design achieves an average speedup of 1.91$\times$ with an area overhead of 0.19\%.
    \item We observe that ray traversal exhibits irregular memory access patterns that cannot benefit from traditional GPU memory coalescing mechanisms. We exploit this insight by dynamically regrouping rays into new coherent groups {during hardware traversal}.
    \item We present the first detailed microarchitectural characterization of NVIDIA RTX 5090 ray tracing cores, which suggests that RT core resources may be allocated at the warp granularity, shows that shared memory is highly active during traversal, and that the pipeline width inside the RT core is less wide than the SIMT pipeline. We leverage these insights to inform our design of Ray-by-Ray.
\end{itemize}

% maybe not here
% However, the fundamental characteristics of ray tracing, including its irregular control flow and data access patterns, introduce significant challenges for massively parallel architectures. Each ray may traverse different paths, access distinct regions of the acceleration structure, and terminate at varying depths, leading to severe execution and data divergence that reduces SIMT efficiency and overall hardware utilization.

\section{Background}
% \subsection{Ray Tracing}
Ray tracing is a fundamental rendering technique in computer graphics that simulates the behavior of light to produce realistic images. In ray tracing, rays are cast from the camera into the scene to determine which objects they intersect. For each ray, the closest intersected primitive is identified, and its material and texture properties are used to compute the final pixel color. The any-hit shader can execute on every potential hit primitive and terminate early when an occlusion or transparency condition is met, while the closest-hit shader runs only for the nearest valid intersection and performs the final shading. Shadow rays often rely on lightweight any-hit shaders to stop traversal as soon as a blocker is found, making soft and hard shadows efficient to compute. To simulate optical effects such as reflections, secondary rays are generated at the point of intersection and traced in the reflected direction to determine what is visible along that path. These shader stages work together as part of the ray-tracing pipeline, enabling effects that rasterization cannot accurately reproduce.

A common application of ray tracing is path tracing, which models the full global illumination of a scene. After a ray hits an object, it continues to bounce through the scene, simulating indirect lighting and diffuse inter-reflections. This process repeats until a termination condition is met, such as reaching a maximum bounce depth or exiting the scene. The accumulated color contributions from all bounces are combined to compute the final pixel shading. While path tracing delivers highly realistic results, it is extremely computationally intensive; each ray may traverse multiple bounces, and rays in the same frame often follow entirely different paths, causing divergence and making efficient parallel execution challenging.

% \subsection{Bounding Volume Hierarchy (BVH)}
% Contemporary computer graphics typically represent objects using triangles as the fundamental primitives. A scene can contain millions of triangles, making it impractical to test each ray against every triangle individually. To accelerate this process, scenes are organized into a hierarchical structure known as a Bounding Volume Hierarchy (BVH). Each node in the BVH is an Axis-Aligned Bounding Box (AABB). Because smaller AABBs are fully enclosed within larger ones, if a ray misses a parent AABB, all of its child AABBs can be skipped. This hierarchical organization enables efficient culling during traversal and greatly reduces the number of required intersection tests. 

% To traverse the BVH, the process starts from the root node. A ray, defined by its origin and direction, is mathematically represented as a line and is tested for intersection against the AABBs in the hierarchy. If the ray intersects a node’s bounding box, traversal continues to its child nodes. When the ray reaches a leaf node, it is tested against the actual primitives contained within that leaf to determine whether a hit occurs. The closest intersected triangle, if any, is recorded as the final hit.

Contemporary computer graphics represent objects using triangles as the basic primitives. To accelerate traversal, scenes are organized into a Bounding Volume Hierarchy (BVH), a tree structure that spatially partitions the scene. Each node in the BVH is an Axis-Aligned Bounding Box (AABB) that encloses a subset of the scene geometry. Internal nodes contain AABBs that bound their children, while leaf nodes contain references to the actual triangles. Because smaller AABBs are fully enclosed within larger ones, if a ray misses a parent AABB, all of its child AABBs can be skipped. This hierarchical organization enables efficient culling during traversal and reduces the number of intersection tests.

To traverse the BVH, the process starts from the root node. A ray, defined by its origin and direction, is represented as a line and is tested for intersection against the AABBs in the hierarchy. The ray-AABB intersection test checks whether the ray enters and exits the bounding box within valid parameter ranges along the ray direction. If the ray intersects a node's bounding box, traversal continues to its child nodes, which are pushed onto a traversal stack or queue for processing. This process repeats recursively or iteratively through the tree. When the ray reaches a leaf node, it is tested against the primitives contained within that leaf using ray-triangle intersection tests to determine whether a hit occurs. Each intersection computes the distance along the ray to the hit point, thereafter the closest intersected triangle within the valid ray parameter range is recorded as the final hit. 
% If multiple triangles are hit, only the nearest one is kept to determine the visible surface.

\section{Ray Tracing on GPUs} \label{microbenchmarking}
In this section, we develop a new profiling tool and use the NVIDIA OptiX ray tracing engine to investigate architectural details of how ray tracing is executed on NVIDIA GPUs. Contemporary GPUs include dedicated hardware units to enable real-time ray tracing. Low-level graphics APIs such as Vulkan and DirectX expose a Ray Tracing Pipeline that allows developers to implement ray tracing through programmable shader stages, including \textit{ray generation} (raygen), \textit{closest-hit}, \textit{any-hit}, \textit{miss}, and \textit{intersection} shaders. Within the \textit{raygen} shader, the \textit{traceRay()} function initiates the ray generation and triggers BVH traversal. The other shaders are executed based on the traversal results. While the software-level pipeline is well defined, the underlying hardware implementation of RT cores remains proprietary and publicly undisclosed.
Consequently, research simulators such as Vulkan-Sim~\cite{saed_vulkansim_2022} rely on simplifying assumptions about how traversal, buffering, and scheduling behave inside the RT core. These assumptions are necessary in the absence of public documentation, but may not reflect how real GPUs are actually implemented. This gap motivates our study: to characterize the behavior of contemporary NVIDIA RT cores and understand their true architectural design.

Our new profiling tool is based on the CUDA Profiling Tools Interface (CUPTI)\cite{_nvidia_cuda_} and GPUVision\cite{pan_architecting_2026}. CUPTI is the low-level library used by NVIDIA Nsight tools to collect performance metrics and hardware events from GPUs. This interface allows us to sample performance counters at extremely fine granularity—on the order of tens of cycles. With this fine-grained information, we can observe how different hardware pipelines interact during execution, providing key insights into the design and behavior of the RT cores.

% We mainly used the optixTriangle example from the Optix SDK 9.0.0. The scene consists of only one triangle. We used this because its simple and the behavior is easily predictable. 
% Within the Optix \textit{raygen} shader, \textit{optixTrace()} is called to initiate the pipeline. It generates the ray, performs traversal and calls the programmable shaders as defined. Once the \textit{optixTrace()} returns, the pipeline is done. 
% There is another variant, \textit{optixTraverse()}, which does traversal only and returns. Then in the \textit{RayGen} shader, the programmer calls \textit{optixInvoke} to invoke the shaders. We used this variant so that we can take a timestamp without the shading part, allowing us to profile the ray traversal part only. The \textit{optixTraverse()} is implemented as a template. Within the template, it calls a function with inline PTX. This is the lowest level of detail we can get from Optix and the implementation of the function is hidden underneath the compiler. We added \textbf{start} and \textbf{end} timestamps around this function to minimize interference of other operations done in the \textit{RayGen} shader. 

To conduct this study, we used the \textit{optixTriangle} example from the NVIDIA OptiX SDK 9.0.0 and ran all experiments on an RTX 5090 GPU. {To confirm that our observations are not specific to a single architecture, we reproduced the same experiments on an Ampere RTX 3070 and found that the conclusions hold across both generations. The results we report are drawn from the RTX 5090.}
The scene consists of a single triangle, chosen for its simplicity and predictable behavior. Within the OptiX \textit{raygen} shader, the \textit{optixTrace()} function is called to initiate the ray tracing pipeline. 
It generates the ray, performs BVH traversal, and invokes the programmable shaders as defined by the pipeline configuration. Once \mbox{\textit{optixTrace()}} returns, the entire ray tracing process is complete. 

OptiX also provides a variant, \textit{optixTraverse()}, which performs only the traversal stage and then returns control to the \textit{raygen} shader. The programmer can subsequently call \mbox{\textit{optixInvoke()}} to execute the shading stages. We used this variant to isolate and measure the traversal stage without the effects of shading. 
% The \textit{optixTraverse()} function is implemented as a template that internally calls a function containing inline PTX instructions, which represent the lowest level of detail accessible through the OptiX API since the implementation beneath this call is handled by the compiler. 
We used inline PTX to sample the $\%clock$ special register and record start and end timestamps around this function; representing the lowest level of visibility exposed to OptiX programmers, thus allowing us to accurately capture the traversal latency.

\subsection{On-Chip Buffer Management} \label{warp_buffer_management}
During traversal, RT cores perform ray intersection tests using ray properties such as $origin$, $direction$, and $t_{max}$. They also track per-thread status information, including whether a thread is active and its current closest hit~\cite{saed_vulkansim_2022,muthler_techniques_traversing_2022,zhu_rtnn_accelerating_2022}. Each ray operates independently, and its data or properties are not shared with other rays within the same warp. In principle, only the active threads in a warp should require buffer space to store this runtime information.

To empirically determine the allocation granularity of RT core resources (i.e., per-warp versus per-thread), we engineered a targeted microbenchmark based on the \textit{optixTriangle} workload. We modified the kernel to selectively mask active threads within a warp and precisely measured the latency of the traversal subroutine. To isolate the RT core pipeline and eliminate execution noise from the broader memory hierarchy, we restricted the ray tracing workload to a single SM, intentionally exiting warps on all other SMs before they could invoke traversal.

% Figure \ref{fig:end_cycle_lane_active} shows start and end cycles of these warps running on the same SM. X-axis is warp index which is sorted by end cycle for better illustration, instead of warp launch order ID. For each warp, we control how many threads are active (doing traversal) and we show two cases, one thread active and all threads active. The vertical span between two lines is the traversal latency of the warp. The block size is 128 threads, and occupancy is 4 thread blocks per SM which is 512 threads, or 16 warps per SM. In both cases we can see that for the first batch consists of 12 warps. In the one thread/warp active case, despite there are more warps launched to the SM and waiting to be traversed, and because only one thread is active, only 12 warps are traversed in parallel, same to the all threads active case. This finding shows inefficiency in RT core resource management that addtional warps cannot utilize the empty slots created by divergence in the warp. 

\begin{figure}[h]
    \centering
    % \vspace{-10pt}
    \includegraphics[width=\linewidth]{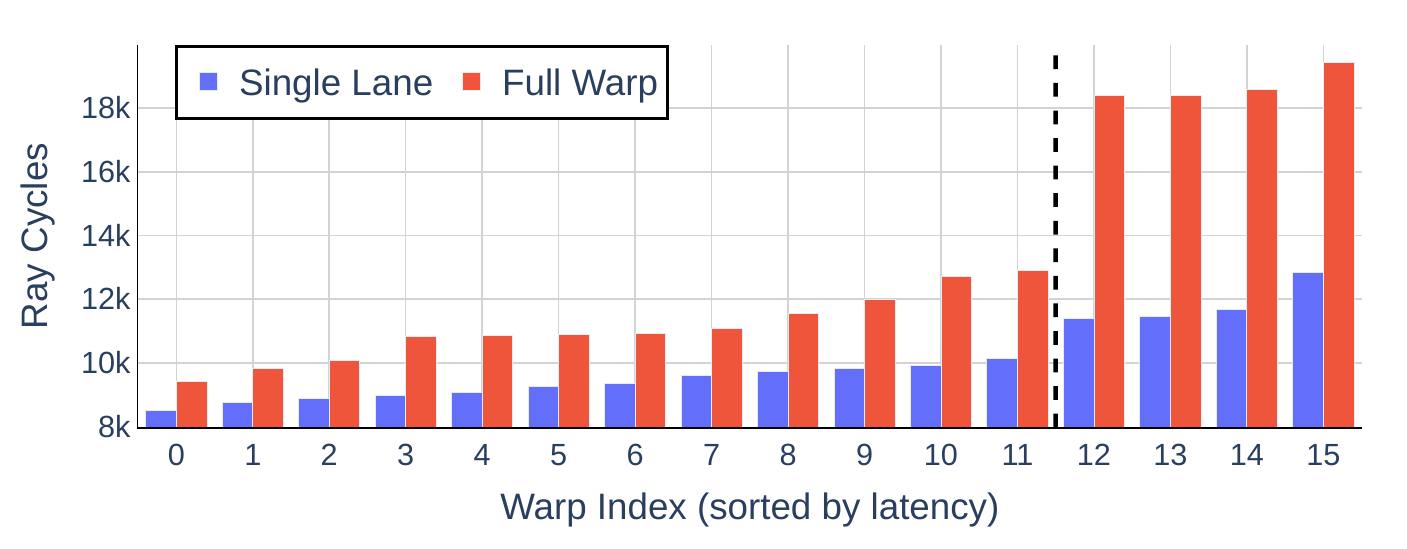}
    % or figure* + \pagewidth for bigger images
    \caption{Measured warp traversal latency under single-lane and full-warp execution. The maximum concurrent occupancy remains constrained to 12 warps in both scenarios, demonstrating that inactive lanes statically reserve on-chip resources and cause severe under-utilization.}
    \label{fig:end_cycle_lane_active}
\end{figure}

Fig.~\ref{fig:end_cycle_lane_active} illustrates the measured warp traversal latency. {We leveraged SER-variant APIs to measure only the latency of the traversal stage.} Our application configuration uses a block size of 128 threads with an SM occupancy of four blocks, yielding 16 active warps per wave of RT execution. We evaluate two distinct scenarios: a \textit{Full Warp} case (all 32 threads active) and a \textit{Single Lane} case (only 1 thread active per warp). By capturing timestamps immediately before and after the traversal instruction, we observe that all 16 warps issue the traversal operation simultaneously.
However, the completion times exhibit a distinct bifurcation: 12 warps complete significantly earlier, while the remaining four experience elevated latency. This delayed execution indicates that the SM's warp scheduler stalls further instruction issue due to a structural hazard at the RT core, revealing a maximum concurrent RT occupancy of exactly 12 warps.
Critically, this 12-warp capacity limit persists in both the \textit{Full Warp} and \textit{Single Lane} scenarios. If the RT core dynamically allocated resources strictly to active threads, the \textit{Single Lane} case, requiring only a fraction of the execution state per warp, should comfortably allow all 16 warps to traverse concurrently. The fact that the bottleneck remains fixed at 12 warps demonstrates that the RT core allocates internal buffer resources at a rigid warp granularity. Consequently, inactive lanes within a divergent warp needlessly consume critical on-chip resources, exposing an inefficiency in contemporary RT core design that restricts overall concurrency.

\begin{tcolorbox}[
    colback=white!15,     % Background color
    colframe=black,      % Border color
    arc=3mm,             % Rounded corners
    boxrule=0.5pt,       % Border thickness
    left=6pt, right=6pt, top=1pt, bottom=1pt, % Padding
    width=\linewidth,    % Full width
]
\textbf{Takeaway \#1:} \textit{RT core resources appear to be allocated on a per-warp basis. When a warp has only a few active threads, the remaining inactive lanes still occupy resources, leading to inefficiency during traversal.}
\end{tcolorbox}

\subsection{Shared Memory Management} \label{stack_management}
% Contemporary GPUs use a short traversal stack to keep track of which BVH nodes to visit next~\cite{vaidyanathan_wide_2022,ylitie_efficient_2017,lee_sgrt_2013,amd_rdna4_2025, jung_hierarchical_2025}. Each stack entry stores the memory address of a node. During traversal, starting from the root node, a depth-first search is performed. When a hit occurs, one child node is visited next while the remaining child nodes are pushed onto the stack for later traversal. If a node is missed, traversal can return early without visiting its children. Once traversal reaches a leaf node or determines that no child nodes need to be visited, the process continues by popping the top of the stack. This repeats until the stack becomes empty. 
Shared memory is widely used in contemporary GPUs to support high-performance graphics and compute workloads. It provides fast, programmer-managed storage that enables efficient data reuse and coordination among threads. Understanding how on-chip memories are utilized in ray tracing workloads is therefore important for characterizing the behavior of RT hardware.

Modern ray tracing systems typically employ a short traversal stack to keep track of which BVH nodes should be visited next~\cite{vaidyanathan_wide_bvh_2022,ylitie_efficient_incoherent_2017,lee_sgrt_mobile_2013,amd_rdna4_instruction_2025,jung_hierarchical_2025}. Each stack entry stores the memory address of a node. Traversal begins at the root node and proceeds as a depth-first search. When a ray intersects a node, one child node is visited immediately while the remaining child nodes are pushed onto the stack for later traversal. If a node is missed, traversal can skip its children entirely. When traversal reaches a leaf node or determines that no additional child nodes need to be visited, execution resumes by popping the next node from the stack. This process repeats until the stack becomes empty.
% The traversal stack is typically small and referred to as a short stack~\cite{vaidyanathan_wide_2022, jung_hierarchical_2025,amd_rdna4_2025}.

\begin{figure}[h]
    \centering
    \includegraphics[width=\linewidth]{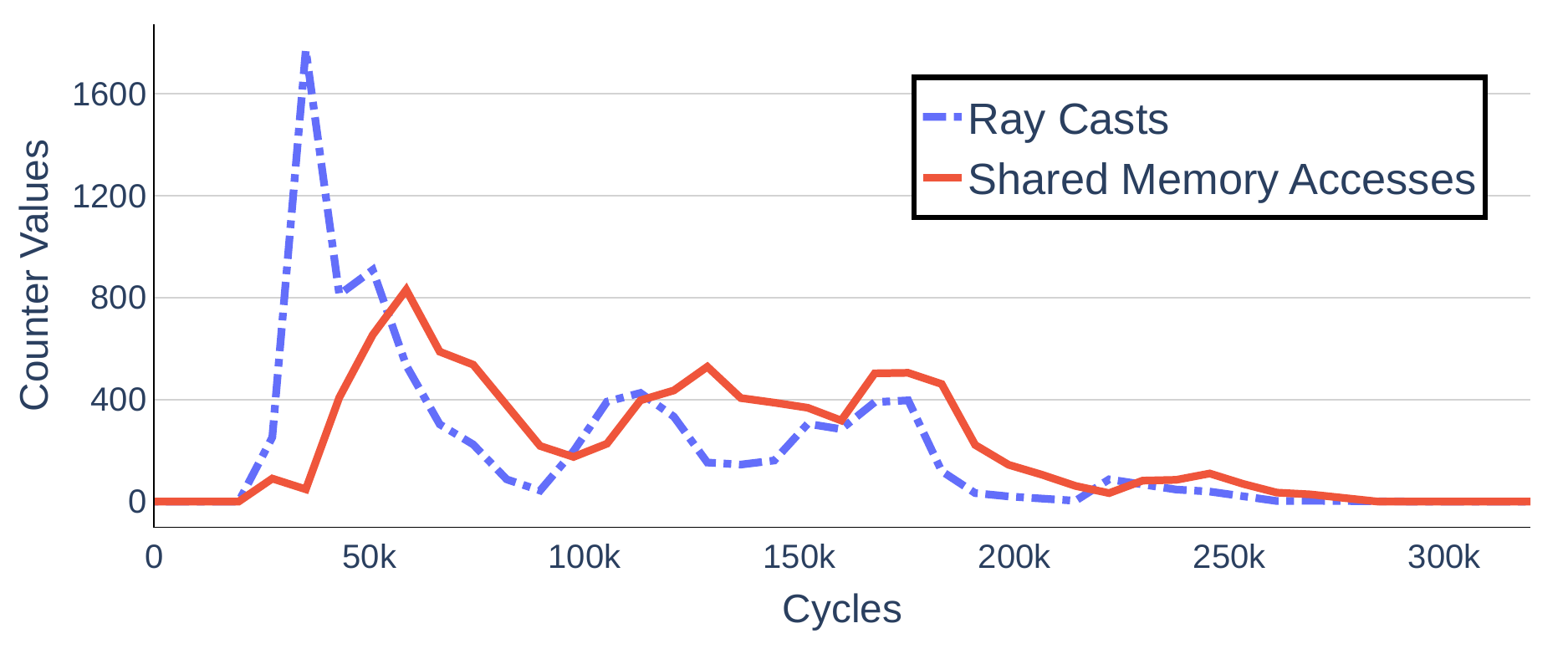}
    % or figure* + \pagewidth for bigger images
    \caption{Shared memory activity profiled alongside RT core ray-cast operations. The strong temporal correlation between these metrics indicates that the RT core utilizes shared memory to manage traversal state.}
    \label{fig:traversal_lds_activity}
\end{figure}

To better understand how NVIDIA's dedicated RT cores manage traversal state, we profile the \textit{optixHair} application using our CUPTI-based profiling tool. We select this workload because its complex geometry induces frequent BVH traversal and therefore sustained RT-core activity.

Fig.~\ref{fig:traversal_lds_activity} shows two metrics sampled during execution: shared memory accesses and RT core ray-cast operations. The two signals exhibit a strong temporal correlation, indicating that traversal-related operations interact closely with shared memory resources. Further evidence from Nsight Compute's Launch Statistics on the RTX~5090 shows that each thread block allocates 8192 bytes of shared memory for 128 threads, corresponding to 64B per thread.

This allocation suggests that a nontrivial amount of on-chip storage is provisioned per ray during traversal. If interpreted in terms of a traversal stack, a 64B allocation could accommodate several stack entries when using 64-bit node addresses (8B per entry), consistent with the short traversal stacks commonly described in prior ray tracing systems~\cite{saed_vulkansim_2022,vaidyanathan_wide_bvh_2022}. While shared memory may support multiple purposes in the traversal pipeline, these observations indicate that traversal-related state can consume a meaningful fraction of available on-chip storage.

From an architectural perspective, this highlights the importance of carefully managing shared memory resources in ray tracing pipelines. Shared memory is heavily utilized across both graphics and compute workloads, and allocating portions of it for traversal state introduces trade-offs with other uses. Designers of future RT hardware must, therefore, consider how traversal data structures interact with shared memory capacity, access frequency, allocation granularity, and bank conflicts when optimizing for both performance and area efficiency.

\begin{tcolorbox}[
    colback=white!15,     % Background color
    colframe=black,      % Border color
    arc=3mm,             % Rounded corners
    boxrule=0.5pt,       % Border thickness
    left=6pt, right=6pt, top=1pt, bottom=1pt, % Padding
    width=\linewidth,    % Full width
]
\textbf{Takeaway \#2:} \textit{Traversal BVH activity strongly correlates with shared memory traffic, suggesting that traversal state is managed through a combination of RT core resources and shared memory.}
\end{tcolorbox}

\subsection{RT Cores Width} \label{rt_core_width}
\begin{figure}[h]
    \centering
    \includegraphics[width=\linewidth]{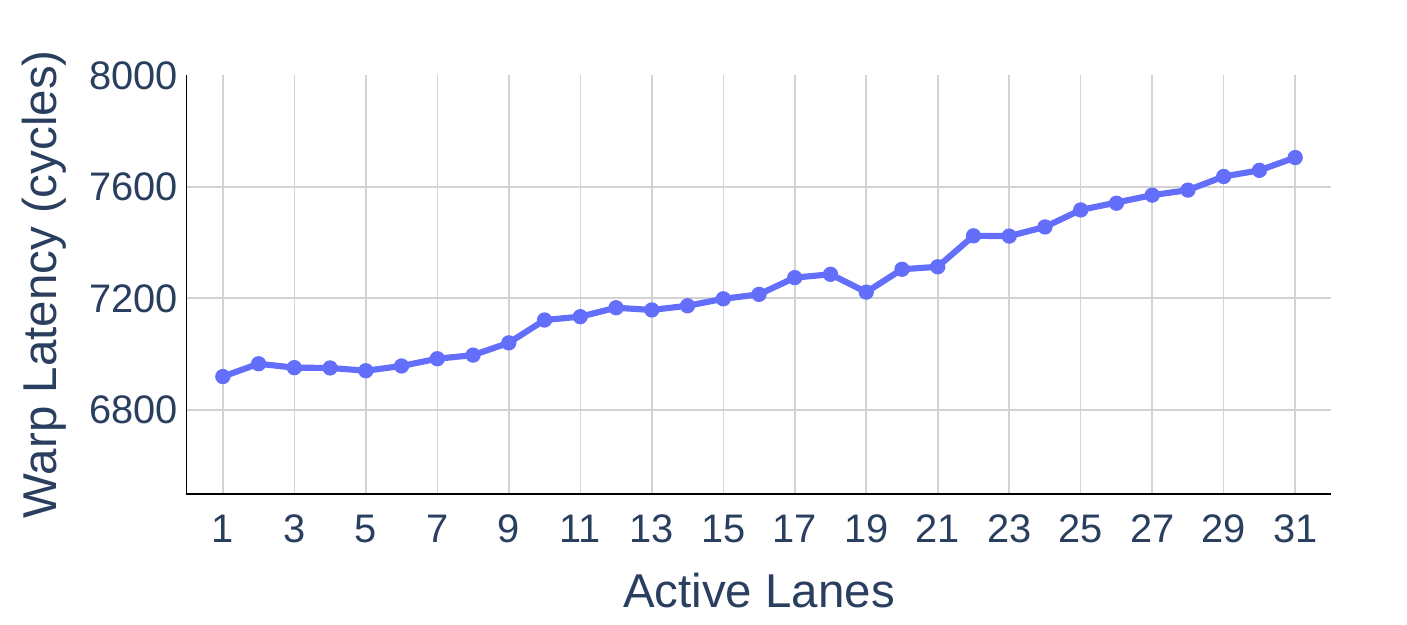}
    % or figure* + \pagewidth for bigger images
    \caption{Measured warp latency as a function of active thread count. The proportional increase in latency for identical, non-divergent rays demonstrates that the RT core employs a serialized, temporal SIMT pipeline.}
    \label{fig:sweep_lane_active}
\end{figure}

GPUs follow a SIMT execution model, where 32 threads are grouped into a warp and execute the same instruction in lockstep. AMD's patent~\cite{saleh_texture_processor_2021} and several prior studies~\cite{shen_rayflex_opensource_2025, barnes_extending_gpu_2024,govindaraju_multicore_architecture_2008,seiler_larrabee_manycore_2008,spjut_trax_multicore_2009,kopta_efficient_mimd_2010,lee_sgrt_mobile_2013} suggest that the RT core may be implemented as a scalar pipeline, where threads within a warp are serialized to perform intersection tests. Other works~\cite{saed_vulkansim_2022, jung_hierarchical_2025, ha_generalizing_ray_2024, chou_treelet_accelerated_2025, tozlu_cooprt_2025,reshetov_multilevel_ray_2005,bigler_design_parallel_2006,wald_interactive_rendering_2001} have instead assumed that the RT core operates as a SIMD pipeline, similar to conventional GPU compute pipelines.
To verify these assumptions, we constructed an experiment using the \textit{optixTriangle} application by launching a single warp and controlling how many threads within the warp invoke \textit{traceRay()}. All rays were configured to be identical, ensuring that memory accesses are perfectly coalesced and the BVH traversal paths are identical, thereby eliminating any divergence. The \textit{traceRay()} call is an API function that triggers a sequence of operations: first, a subroutine sets up the ray data and prepares the RT core, then an RT core instruction is issued to initiate traversal in hardware. 
Fig.~\ref{fig:sweep_lane_active} shows the results, where the y-axis represents latency in cycles and the x-axis represents the number of active threads per warp. The measured latency includes the overhead of the entire \textit{traceRay()} API function, the subroutine setup, and the RT core instruction execution. Since all rays are identical, each thread should experience the same execution path through the subroutine and RT core instruction. However, we observe that latency continues to increase as more threads become active. The rest of the GPU utilizes conventional SIMT pipelines, where all threads execute the same operations simultaneously. Since we deliberately removed memory divergence in our application, the extra latency comes directly from the RT core. This behavior suggests that NVIDIA's RT core employs a temporal SIMT style pipeline \cite{keckler_gpus_2011}, where the warp's threads are serialized onto a narrower execution datapath and executed over multiple cycles.

From a modeling perspective, treating RT cores as SIMD pipelines remains valid. Although latency increases as more lanes invoke \textit{traceRay()}, BVH traversal is highly memory-intensive, and its latency is largely dominated by memory operations. Consequently, the additional delay from serializing threads within a warp is often hidden by memory latency. The behavior can still be represented as a single SIMD operation with higher latency.

% However, this finding does not invalidate prior work that assumed SIMD-like RT cores. Although latency may increase as more lanes invoke \textit{traceRay()}, BVH traversal is highly memory-intensive, and its latency is largely dominated by memory operations. Consequently, the additional delay from serializing threads within a warp is often hidden by memory latency. From a modeling perspective, the behavior can still be represented as a single SIMD operation with higher latency.

% One interesting observation is that when the warp is fully coherent with all 32 threads active, traversal latency is lowest. Figure~\ref{fig:banks_read_31_32} shows the number of L1 cache bank read accesses for two runs: one with 31 active threads and one with 32 active threads. When all threads are active, the number of accesses decreases. This indicates that the RT core performs memory coalescing, similar to that described for the Fermi architecture in PTX 1.0~\cite{nvidia_nvidia_cuda_2007}.

% \begin{figure}[h]
%     \centering
%     \includegraphics[width=\linewidth]{figs/lds_31_32.pdf}
%     % or figure* + \pagewidth for bigger images
%     \caption{LD/ST Unit requests during ray tracing traversal with 31 versus 32 active threads per warp. Fully active warps exhibit reduced memory accesses.}
%     \label{fig:banks_read_31_32}
% \end{figure}

% While the exact reason remains unclear, we suspect that certain forms of control divergence may be mitigated within the RT core under this condition.

\begin{tcolorbox}[
    colback=white!15,     % Background color
    colframe=black,      % Border color
    arc=3mm,             % Rounded corners
    boxrule=0.5pt,       % Border thickness
    left=6pt, right=6pt, top=1pt, bottom=1pt, % Padding
    width=\linewidth,    % Full width
]
\textbf{Takeaway \#3:} \textit{The RT core is designed as a temporal SIMT pipeline, where threads within a warp are serialized for processing.}
\end{tcolorbox}

\section{Ray-by-Ray} \label{design}

In this section, we begin by introducing the baseline RT unit model used in our study, which reflects the typical design principles of contemporary GPU ray tracing hardware. Building on the characterization results discussed in Section~\ref{microbenchmarking}, we analyze the key inefficiencies observed in the baseline design, focusing on how resource allocation and warp-level management impact overall traversal efficiency. These insights motivate the need for a more flexible and fine-grained approach to managing hardware resources during ray traversal. Finally, we present our proposed design Ray-by-Ray, which introduces thread-level resource allocation mechanisms to improve utilization and enhance overall ray tracing performance.

\begin{figure*}
    \centering
    \includegraphics[width=\textwidth]{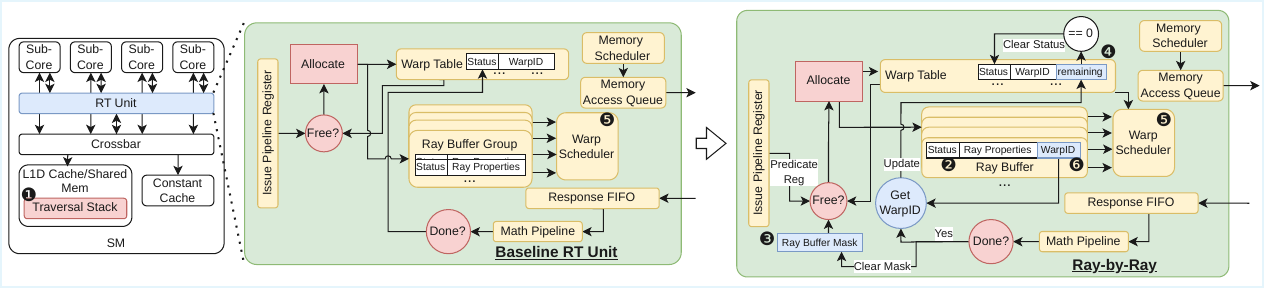}
    % or figure* + \pagewidth for bigger images
    \caption{Comparison of the baseline RT unit design and Ray-by-Ray with ray-granularity resource management.}
    \label{fig:rt_core_design}
\end{figure*}

\subsection{Baseline RT Unit}
% A warp starts traversal by calling the $trace\_ray()$ instruction. Each thread of the warp is just one ray. Similar to traditional SIMD pipelines, the SM warp scheduler tests the scoreboard for dependencies. Once the dependencies are resolved and the warp is eligible, the warp is issued to the issue register. If there are empty warp slots within the RT unit, it accepts the warp and moves the warp from the pipeline register to the internal warp buffer to start traversal. Each cycle, the unit checks every warp and a warp is selected and memory requests from active threads in the warp are scheduled to be issued. Also each cycle, the Response FIFO is checked to see if any previously issued loads are back. If true, then the data is forwarded to the ray test units and performs the math to do the ray-box test for internal box nodes, or the ray-triangle test for leaf nodes (also transformation for transformation nodes). The number of active threads per warp is tracked in the warp buffer. The unit manages the stack according to keep track what nodes need to be tested. Once the stack is empty, the traversal is done and the thread is marked. The thread writes the traversal result to the memory and is later used by the programmable shaders to perform shading. Once all threads within a warp are done, the warp commits and returns, and the subsequent instructions depending on this can proceed. 

A warp begins traversal by invoking the $trace\_ray()$ instruction, where each thread in the warp corresponds to one ray. Similar to traditional SIMD pipelines, the SM warp scheduler inside each sub-core checks the scoreboard for dependencies. The sub-core warp scheduler arbitrates among all warps within the sub-core to determine if the warp can be issued. Once all dependencies are resolved and the warp becomes eligible for execution, it is issued if there are available warp table slots within the RT unit. 

As shown in Fig.~\ref{fig:rt_core_design}, each cycle, the RT unit scheduler \text{\ding{186}} selects a warp and schedules memory requests from its active threads. The Response FIFO is also checked each cycle for completed loads. When data returns, it is forwarded to the ray test units, which perform ray-box tests for internal nodes, ray-triangle tests for leaf nodes, and transformations for transformation nodes. The unit tracks the number of active threads per warp in the warp table and manages the short stack \text{\ding{182}} to record which nodes still need testing.
When the stack becomes empty, traversal completes and the thread is marked as done. The thread writes its traversal result to memory for later use by programmable shaders during shading. Once all threads in a warp finish, the warp commits and returns, allowing dependent instructions to proceed.

% The first source arises when warps are not fully occupied during ray construction. In path tracing (PT) workloads, after a ray completes one traversal, a new ray is generated at the intersection point to simulate the physical reflection or scattering of light within the scene. This process repeats iteratively until a predefined bounce limit is reached or the ray exits the scene. As a result, even if a warp begins fully populated, rays terminate at different bounce counts, leaving inactive threads in subsequent traversals and reducing overall RT efficiency. In the worst case, only a single thread in the warp remains active and may continue bouncing several times before termination. This significantly lowers traversal efficiency, as only a small number of threads are processed in parallel. Although Section~\ref{rt_core_width} shows that the RT core operates in a MIMD-like manner than running less threads active should be faster than running a rull warp in theory, traversal remains highly memory-latency-bound and requires a sufficient number of active rays to exploit memory-level parallelism (MLP) and effectively hide latency. 

\begin{figure}
    \centering
    \includegraphics[width=\linewidth]{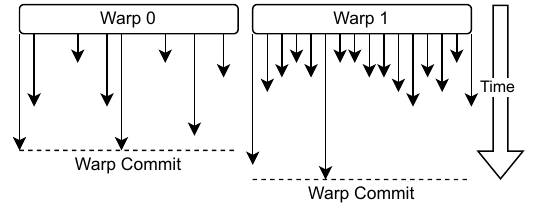}
    % or figure* + \pagewidth for bigger images
    \caption{Illustration of two warp-granularity resource management inefficiencies. Warp 0: reduced parallelism when only a subset of threads remain active. Warp 1: tail effect where completed threads hold resources while waiting for the slowest thread in the warp to finish traversal.}
    % \vspace{-10pt}
    \label{fig:divergent_ineff}
\end{figure}

We identify two major sources of efficiency loss in the baseline RT core traversal.
% claude:
The first source arises when warps are not fully occupied during ray construction. For example, in path tracing workloads, after a ray completes one traversal, a new ray is generated at the intersection point to simulate the physical reflection or scattering of light within the scene. This process repeats until a predefined bounce limit is reached or the ray exits the scene. As a result, even if a warp begins fully populated, rays terminate at different bounce counts, leaving inactive threads occupying resources in subsequent traversals, thus reducing overall RT efficiency. In the worst case, only a single thread in the warp remains active and may continue bouncing several times before termination. As illustrated by \textit{warp 0} in Fig.~\ref{fig:divergent_ineff}, this lowers traversal efficiency, as only a small number of threads are processed in parallel. 
Although Section~\ref{rt_core_width} shows that the RT core is temporal SIMT, and running fewer active threads should theoretically be faster than running a full warp, traversal remains highly memory latency bound and requires a sufficient number of active rays to exploit memory-level parallelism and hide latency. 

% The second source is that even some threads within the warp finishes traversal earlier than others, they have to wait until the longest thread to finish before commiting. In ray tracing, eay ray is individual and travels completely different paths with other rays. Even though the instruction follows a SIMT manner, the work performed by threads within the warp are not uniform, which is not the same as a traditional SIMT instruction such as $FMA$ where all threads are doing exactly one Fused multiply-add and return. This causes significantly tailing effect where several rays finish much later than the others. Even though the warp may enter the RT core with all lanes active, most of the time it's possible that only a few threads are active which degrades RT efficiency. As shown in \ref{warp_buffer_management}, the RT core buffers and the traversal stack is managed on warp level which the resources can only be released after all threads in the warp is finished. While the warp must commit traversal together because the rest of the program follows the SIMT programming model, the resources to store values that are needed during traversal can be freed as soon as the corresponding thread has finished traversal. 

The second source of inefficiency occurs when some threads within a warp complete traversal earlier than others, but must wait for the longest-running thread to finish before the warp can commit. In ray tracing, each ray is independent and may traverse completely different paths. Although the instruction execution follows the SIMT model, the work performed by threads within a warp is highly non-uniform, unlike traditional SIMT instructions, where all threads perform one operation simultaneously. This leads to a pronounced tail effect, where a few rays continue running long after others have completed. As illustrated by \textit{warp 1} in Fig.~\ref{fig:divergent_ineff}, even if a warp enters the RT core with all lanes active, it is common for only a small number of threads to remain active during later stages of traversal, reducing RT efficiency. While the warp must commit traversal collectively to maintain the SIMT programming model, per-thread resources used during traversal could, in principle, be freed as soon as their corresponding threads have completed. However, as shown in Section~\ref{warp_buffer_management}, the RT core buffers and traversal stacks \text{\ding{182}} are managed at the warp level, meaning that resources cannot be released until all threads in the warp have finished, leading to inefficient resource utilization.

\subsection{Fine-Grained Ray-based Allocation}
Given these observations, we propose a fine-grained resource management mechanism that allocates and releases RT unit resources at the thread level rather than at the warp level. This design addresses both major sources of inefficiency observed in RT traversal. When a warp issues an RT instruction to initiate traversal, only the active threads are allocated the necessary resources, whereas inactive threads consume no additional storage. As each thread completes traversal, its allocated resources are immediately released, allowing new threads or warps to utilize them. Once sufficient resources become available, a pending warp can be issued to the RT unit to begin traversal. By dynamically allocating resources at per-thread granularity, the design improves RT unit utilization under divergent workloads and increases the effective number of rays in flight, leading to higher overall throughput. 

\begin{figure}[h]
    \centering
    \includegraphics[width=\linewidth]{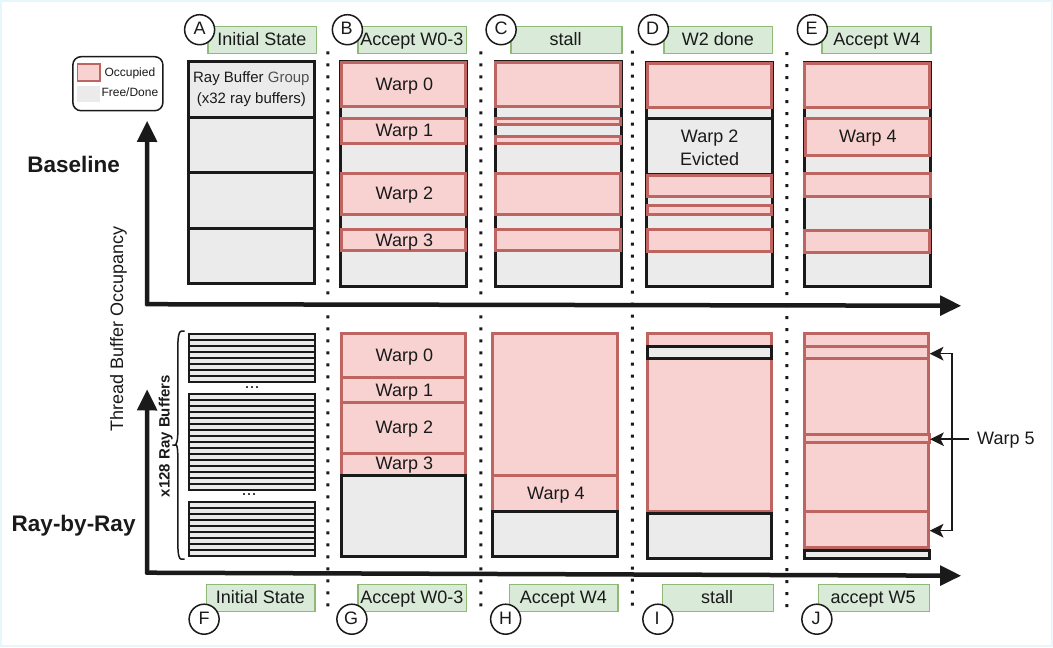}
    % or figure* + \pagewidth for bigger images
    \caption{Comparison between baseline and proposed buffer management. The top is the baseline, where buffer groups are managed at the warp level. Bottom is Ray-by-Ray, allowing fine-grained per-ray buffer management.}
    \label{fig:timeline_occupancy}
\end{figure}

This design is possible because, as shown in Section~\ref{rt_core_width}, the RT unit does not execute 32 lanes in parallel. Consequently, rays do not need to originate from the same warp. In a traditional SIMD pipeline, repacking lanes requires lane migration, which incurs significant overhead. This is not an issue for the RT unit: ray state is stored in dedicated RT-unit structures rather than in the vector register file, and traversal is already serialized. 
Fig.~\ref{fig:timeline_occupancy} illustrates a timeline comparison demonstrating how our proposed fine-grained resource management addresses the inefficiencies of the baseline. Both configurations provide on-chip buffering for 128 rays. However, the baseline manages these resources in four discrete warp-level groups \textcircled{\scriptsize{A}}, whereas our design utilizes a flattened 128-entry pool \textcircled{\scriptsize{F}}. Under the baseline, warp-granular allocation causes internal fragmentation when warps are partially occupied \textcircled{\scriptsize{B}}. Subsequent workloads such as warp 4 are blocked from entering the RT unit \textcircled{\scriptsize{C}} despite the presence of sufficient free buffer entries. These entries are only released once the entire warp commits collectively \textcircled{\scriptsize{D}}, at which point, warp 4 is finally accepted \textcircled{\scriptsize{E}}.
In contrast, the proposed design allocates and manages ray buffers individually. When warps 0 to 3 are dispatched, only active threads consume buffer slots \textcircled{\scriptsize{G}}, which prevents fragmentation and allows warp 4 to be accepted immediately \textcircled{\scriptsize{H}}. Furthermore, our mechanism reclaims resources at ray granularity as soon as a ray completes its traversal \textcircled{\scriptsize{I}}, removing the constraint for threads to wait for the longest-running ray in a warp to finish before releasing on-chip state. Once sufficient buffers become available, warp 5 is accepted \textcircled{\scriptsize{J}}, and its threads are scattered within the flattened 128-entry pool to maximize RT unit utilization and hide memory latency. 

\subsubsection{Buffer Management} \label{buffer_management}
As detailed in Section~\ref{warp_buffer_management}, contemporary RT units manage on-chip buffers at warp granularity. This allocation is inefficient because inactive lanes continue to occupy buffer entries without performing work. To address this, we reorganize the resource allocation as shown in Fig.~\ref{fig:rt_core_design}. Instead of maintaining separate ray buffer groups, we implement a unified ray buffer \text{\ding{183}} with the same total capacity. When a warp is dispatched, each active thread is assigned to an individual ray buffer entry instead of binding the entire warp. In the baseline architecture, a direct mapping between warps and buffer groups ensures that a free warp table entry guarantees a free resource group. Our design decouples this relationship, requiring the issue logic to verify both warp table and ray buffer availability. This check is performed by evaluating the warp predication mask against a new Ray Buffer Mask \text{\ding{184}}, which uses one bit per ray buffer entry. A warp is only accepted from the issue pipeline register if a warp table entry is available and the unified pool contains sufficient ray buffers for all active threads.
Once a warp is admitted to the RT unit, the hardware assigns each active thread to an available ray buffer entry. Although parallel allocation logic could enable single-cycle initiation, the implementation would require substantial hardware overhead. We therefore utilize a serialized allocation mechanism that processes one thread per cycle. This approach introduces an initiation interval of one cycle per active lane, amounting to 32 cycles for a fully active warp. This latency is negligible relative to ray traversal, which typically spans thousands to tens of thousands of cycles.

% In the baseline architecture, the RT unit scheduler selects one warp per cycle to issue memory requests for BVH traversal. Our fine-grained ray buffer organization replaces explicit warp tracking with a unified resource pool. The scheduler instead issues groups of 32 rays from contiguous buffer entries, which preserves the baseline coalescing width and memory throughput without increasing area overhead. This mechanism decouples ray execution from the original warp assignment by dynamically forming ray groups based on buffer occupancy. This approach allows the RT unit to maintain high utilization and throughput even when rays within the original warps exhibit significant execution divergence.

In the baseline architecture, the RT unit scheduler selects one warp per cycle to issue memory requests for BVH traversal. Our fine-grained ray buffer organization replaces explicit warp tracking with a unified resource pool. The scheduler instead issues groups of 32 rays from contiguous buffer entries, which preserves the baseline coalescing width and memory throughput without increasing area overhead. This mechanism decouples ray execution from the original warp assignment by dynamically forming ray groups based on buffer occupancy. Although rays from the same warp may scatter across non-contiguous entries, the RT unit maintains peak utilization without fragmentation by processing rays in buffer order. This approach allows the unit to better hide memory latency through increased parallelism under significant execution divergence.

% By decoupling ray allocation from warp structure, the proposed design allocates ray buffer entries only to active lanes, eliminating resource waste on inactive threads. While threads from the same warp may occupy non-contiguous buffer entries, this poses no fragmentation concern. By processing rays in buffer order rather than original warp order, the RT unit maintains efficient resource utilization without fragmentation.

As described in Section~\ref{stack_management}, the traversal stack \text{\ding{182}} resides in shared memory. In the baseline, each thread owns a chunk of shared memory for its stack, and reordering threads produces dynamic, unstructured bank-access patterns that can lead to bank conflicts. To avoid this, we decouple the stack from the thread and instead attach it to each ray-buffer entry. Our key observation is that the stack is only needed during traversal, meaning it is required only for the duration of the ray’s lifespan inside the RT unit. Because our design does not preserve the original warp order and traverses rays based on buffer-entry order, shared-memory accesses follow a fixed and structured pattern, eliminating potential bank conflicts. Using fixed shared-memory allocations for fixed-function pipelines is not new in graphics. For example, shared memory is commonly used to store vertex attributes during geometry processing~\cite{pan_crisp_2024, kerbl_revisiting_2018, tine_vortex_2021}.

The RT unit tracks resource occupancy through a single-bit Ray Buffer Mask \text{\ding{184}}, which is scanned during dispatch to assign active threads to available slots. While deallocation is fine-grained, allocation maintains warp atomicity to simplify scheduling and tracking logic: a warp is admitted only if the unified pool contains sufficient free entries for all active threads. This constraint prevents partial warp issuance and ensures consistent state management at the scheduling boundary. 

To manage warp retirement without a fixed buffer structure, we implement a per-warp completion counter \text{\ding{185}} initialized to the active thread count at dispatch. As the math pipeline completes ray tests, the RT warp scheduler \text{\ding{186}} checks the traversal stack \text{\ding{182}} to determine ray termination. Upon termination, the unit immediately reclaims the resource by clearing the corresponding bit in the Ray Buffer Mask \text{\ding{184}} and decrementing the per-warp counter \text{\ding{185}}. Because rays are scattered across the unified pool, each buffer entry includes a Warp ID field \text{\ding{187}} to identify the originating warp. Once the remaining counter reaches zero, the warp commits the $trace\_ray()$ instruction and the warp table entry is cleared. This decoupled management enables the hardware to track a larger pool of concurrent warps, increasing memory-level parallelism and effectively hiding traversal latency through improved resource utilization.

\subsubsection{Memory Coalescing} \label{coalescing}
% In ray traversal, rays within the same warp operate independently. 
In GPGPU kernels, warp reordering often disrupts established memory access patterns. While reordering improves SIMT efficiency, the dynamic reassignment of threads across warps compromises the coalescing optimizations typical in programmer-written kernels. This trade-off is not a concern for RT units. Our key observation is that although rays exhibit spatial locality from screen-space tiling, ray tracing memory access patterns offer negligible opportunities for coalescing. Contemporary GPU architectures employ wide Bounding Volume Hierarchy (BVH) structures such as BVH4 and BVH8, where a BVH8 node typically occupies 128~bytes~\cite{lier_cpustyle_simd_2018,ernst_multi_bounding_2008,wald_getting_rid_2008,meister_survey_bounding_2021,grauer_minimizing_ray_2025}. At this granularity, a single BVH node fetch initiates a 128~B load, fully utilizing an L1 cache line and inherently avoiding memory divergence.

In our proposed architecture, rays map to arbitrary buffer entries rather than maintaining strict warp order. Although such reordering would typically increase memory divergence in traditional SIMT execution \cite{fung_dynamic_2007}, ray traversal is inherently divergent. While rays may exhibit perfect coalescing at the BVH root, they rapidly diverge after only a few intersection tests, diminishing the sustained benefit of traditional coalescing.

Coalescing still occurs when adjacent rays access data within the same cache line, but grouping is no longer constrained by the original warp assignment. This reordering may result in a marginal increase in the number of issued load instructions, as previously coalesced rays may be distributed across different scheduler groups. Crucially, this increase in load requests does not translate to higher L2 memory traffic. Contemporary GPUs utilize Miss Status Holding Registers (MSHRs) to track outstanding misses and merge concurrent requests targeting the same cache line \cite{dublish_evaluating_mitigating_2017}. When multiple rays request a pending address, the MSHR merges these requests into a single entry, preventing redundant memory transactions.

The volume of concurrent memory requests is strictly constrained by the pointer-chasing nature of BVH traversal. Because the address of a subsequent child node is only resolved after the current node is fetched and the intersection test is completed, each ray follows a serial dependency chain that permits only one outstanding memory miss at a time. This traversal dependency prevents the accumulation of excessive in-flight requests. Modern GPU architectures provide sufficient MSHR capacity \cite{khairy_accelsim_2020, khairy_exploring_modern_2020} to accommodate the additional requests generated by our design without reaching structural limits or degrading throughput.

\section{Methodology} \label{methodology}
\begin{figure*}[]
    \centering
    \includegraphics[width=\linewidth]{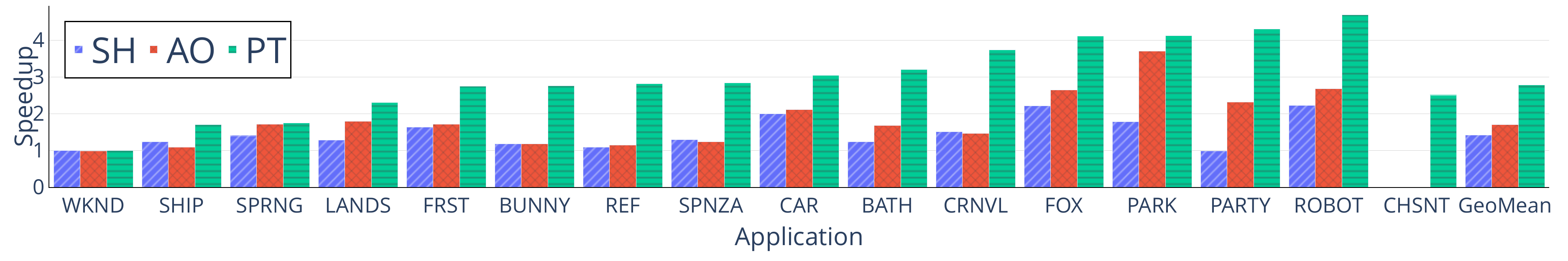}
    % or figure* + \pagewidth for bigger images
    \caption{Normalized speedup of Ray-by-Ray. Blue bars represent Shadow Rays (SH), red bars represent Ambient Occlusion (AO), and green bars represent Path Tracing (PT). The geometric mean speedup across all workloads is 1.91$\times$.}
    \label{fig:speedup}
\end{figure*}
We implement our proposed design in Vulkan-Sim v2.0~\cite{saed_vulkansim_2022,bakhoda_analyzing_2009,ariel_visualizing_2010}, a cycle-level GPGPU simulator that integrates Vulkan with Mesa's frontend. Vulkan-Sim models the complete GPU pipeline, including shader cores, memory hierarchy, and ray tracing units, allowing us to evaluate microarchitectural modifications with cycle-level timing. We extend the simulator's RT unit model to support our ray buffer-based resource management mechanism.

\begin{table}[htbp]
\caption{LumiBench Benchmark Scenes}
\label{tab:lumibench}
\resizebox{\columnwidth}{!}{%
\begin{tabular}{|c|c|c|c|c|c|}
\hline
\textbf{Scene} & {\textbf{\begin{tabular}[c]{@{}c@{}}BVH \\ Depth\end{tabular}}} & {\textbf{\begin{tabular}[c]{@{}c@{}}BVH \\ Size (MB)\end{tabular}}} & \textbf{Scene} & {\textbf{\begin{tabular}[c]{@{}c@{}}BVH \\ Depth\end{tabular}}} & {\textbf{\begin{tabular}[c]{@{}c@{}}BVH \\ Size (MB)\end{tabular}}} \\ \hline
WKND           & 7                   & 0.2                    & PARTY          & 14                  & 143.8                  \\ \hline
SHIP           & 12                  & 0.5                    & SPRNG          & 14                  & 164.3                  \\ \hline
BUNNY          & 11                  & 12.2                   & LANDS          & 12                  & 279.2                  \\ \hline
SPNZA          & 16                  & 22                     & FRST           & 14                  & 348.6                  \\ \hline
CHSNT          & 12                  & 25.5                   & FOX            & 15                  & 597.8                  \\ \hline
REF            & 13                  & 37.1                   & PARK           & 14                  & 501.9                  \\ \hline
CRNVL          & 16                  & 37.3                   & CAR            & 16                  & 1233.6                 \\ \hline
BATH           & 16                  & 104.2                  & ROBOT          & 18                  & 1721.3                 \\ \hline
\end{tabular}%
}
\end{table}

Our evaluation utilizes benchmarks from LumiBench \cite{liu_lumibench_2023}, which provides 16 scenes (Table~\ref{tab:lumibench}) spanning path tracing, ambient occlusion, and shadow ray techniques. Because Ray-by-Ray improves performance by enabling more warps to traverse in parallel rather than by accelerating any individual warp, its effectiveness is governed by the warp pressure presented to a single SM, i.e., the number of warps per SM available to refill the RT unit. This quantity is an SM-local property. 
{Within a GPU generation, the SM microarchitecture is fixed; larger GPUs replicate more SMs and scale L2 capacity and memory bandwidth in proportion to SM count. Consequently, reproducing the correct warps-per-SM density on a small configuration faithfully reproduces the per-SM behavior of a much larger GPU \cite{rogers_variable_warp_2015}, and the relevant pairing to match is the GPU configuration and the rendering resolution rather than the absolute SM count.}

The challenge is that the default operating point is far from representative. At LumiBench's default $128 \times 128$ resolution, the 30-SM RTX 2060 configuration shipped with Vulkan-Sim yields only 17 warps per SM, well below the peak hardware occupancy of 32 warps per SM. The GPU is underutilized and never exercises the RT unit under realistic pressure. For reference, an RTX 5090 rendering a 4K scene operates at roughly 1.5k warps per SM. Naively raising the resolution is not viable: even after software-only optimizations that improved simulation speed by 2$\times$, higher resolutions lead to impractical simulation times \cite{liu_lumibench_2023, avalosbaddouh_principal_2021}. To approximate realistic per-SM density within these limits, we instead pair a validated configuration based on the NVIDIA Ampere Jetson Orin \cite{pan_crisp_2024} with an increased $256 \times 256$ resolution, giving a GPU-resolution pairing whose warps-per-SM density is representative of a typical desktop workload while remaining tractable to simulate. Even with this targeted configuration, a complete data-collection cycle requires one week of cluster execution time, which prevents any further increase in workload scale.

\begin{table}[htbp]
\caption{Simulation Configuration}
\label{tab:sim_config}
\resizebox{\columnwidth}{!}{%
\begin{tabular}{|cl|cc|}
\hline
\multicolumn{2}{|c|}{}                              & \multicolumn{1}{c|}{\textbf{Baseline}} & \textbf{Ray-by-Ray} \\ \hline
\multicolumn{2}{|c|}{\# SMs}                        & \multicolumn{2}{c|}{16}                               \\ \hline
\multicolumn{2}{|c|}{\# Registers / SM}             & \multicolumn{2}{c|}{65536}                            \\ \hline
\multicolumn{2}{|c|}{L1 Data Cache + Shared Memory} & \multicolumn{2}{c|}{192 KB, 256 MSHR Entries}  \\ \hline
\multicolumn{2}{|c|}{\# Warps / Sub-Core}           & \multicolumn{2}{c|}{16}                               \\ \hline
\multicolumn{2}{|c|}{\# Sub-Core / SM}              & \multicolumn{2}{c|}{4}                                \\ \hline
\multicolumn{2}{|c|}{\# Exec Units}                 & \multicolumn{2}{c|}{4 FPs, 4 SFUs, 4 INTs, 1 RT Unit} \\ \hline
\multicolumn{2}{|c|}{RT Throughput}                 & \multicolumn{1}{c|}{\;\;\;\;\;\;\textbf{4 Warps}\;\;\;\;\;\;} & \textbf{128 Rays} \\ \hline
\multicolumn{2}{|c|}{L2 Cache}                      & \multicolumn{2}{c|}{4 MB, 32 Banks}                             \\ \hline
\multicolumn{2}{|c|}{Memory}                        & \multicolumn{2}{c|}{LPDDR5, 200 GB/s}                 \\ \hline
\end{tabular}%
}
\end{table}

% To address this, we adopt an NVIDIA Ampere Jetson Orin configuration from CRISP~\cite{pan_crisp_2024}, which models a smaller 16-SM GPU, and increase the resolution to 256 × 256. This setup generates 128 warps per SM. For context, an RTX 5090 with 170 SMs rendering a 4K scene would have a warp-to-SM ratio of approximately 1.5k. Even with our adjusted configuration, the number of warps per SM remains significantly lower than what would occur on real GPUs at realistic resolutions. 

% {rethink about this}
% We conducted the microbenchmarking studies in Section~\ref{microbenchmarking} on both an NVIDIA Blackwell RTX 5090 and an Ampere RTX 3070. Our results show no significant architectural changes between these generations that would affect our evaluation. Since no validated Blackwell architecture configurations are available for simulation, we use a validated Ampere configuration, which accurately captures the relevant architectural details.

\begin{figure}[htbp]
    \centering
    \includegraphics[width=\linewidth]{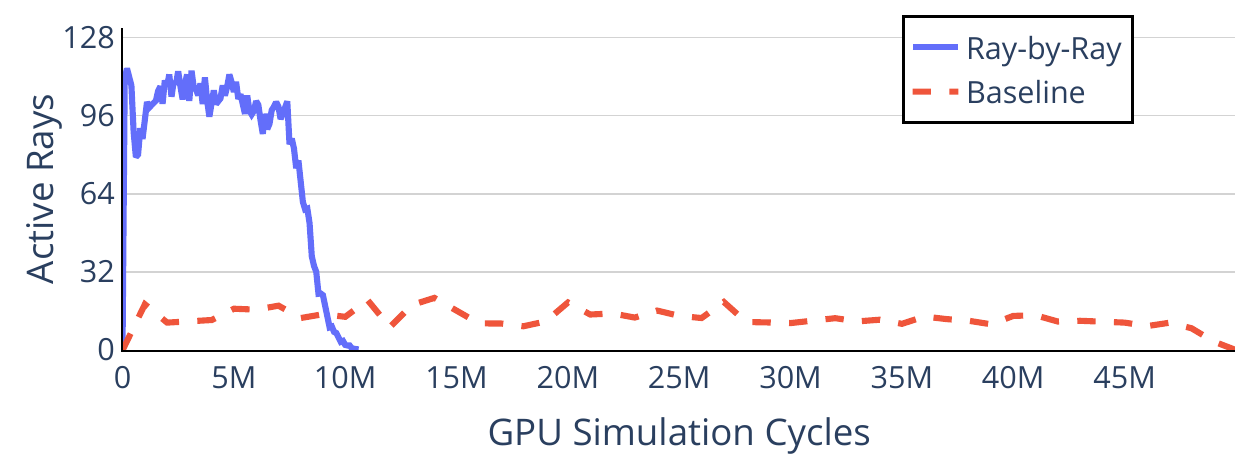}
    % or figure* + \pagewidth for bigger images
    \caption{Average active rays per SM for the ROBOT scene during path tracing. Fine-grained resource management enables more rays to traverse in parallel.}
    \label{fig:robot_active_rays}
    % \vspace{-15pt}
\end{figure}

Table~\ref{tab:sim_config} shows the configuration. In the baseline design, the RT unit traverses 4 warps (32 threads per warp, 1 ray per thread) concurrently. To maintain equivalent throughput, we evaluate the design with 128 ray buffer entries.

\subsection{Performance} 
Fig.~\ref{fig:speedup} shows the normalized speedup of the proposed design. We evaluate each scene with three ray tracing techniques: Shadow Rays (SH), Ambient Occlusion (AO), and Path Tracing (PT). In the default LumiBench suite, CHSNT appears only in the PT category. These techniques achieve speedups of \textbf{1.42$\times$}, \textbf{1.70$\times$}, and \textbf{2.78$\times$} over the baseline, respectively. PT achieves the highest speedup because rays do not terminate immediately after hitting an object. Instead, each ray spawns another ray from the hit point to simulate light transport through the scene. As rays bounce through different paths, the number of active rays within each warp decreases over time, reducing traversal efficiency.

To quantify this effect, we sample counters during simulation to measure active rays within the RT core. Fig.~\ref{fig:robot_active_rays} shows the average active rays for the ROBOT scene with path tracing. In the baseline, only a few rays remain active despite all warp slots being occupied. This demonstrates that rays terminating at different times cause severe divergence, leaving few threads active per warp and reducing RT core utilization. The proposed design addresses this by allocating on-chip RT resources at ray granularity rather than warp granularity. Inactive lanes no longer consume resources, enabling more warps to traverse in parallel. This optimization increases active ray count to nearly the theoretical maximum of 128 rays.

SH and AO exhibit higher ray coherence than PT, yet still show significant speedup. As described in Section~\ref{design}, even when a warp begins traversal with all lanes active, rays traverse different paths through the scene geometry. This creates large variations in completion time, producing a tail effect where only a few rays remain active after several BVH node tests. In the baseline design, on-chip buffers are allocated at warp granularity. Therefore, rays that finish early cannot release their resources until the last thread in the warp completes, reducing RT core efficiency. The proposed design manages resources at ray granularity. Each ray releases its resources immediately upon completion, making them available for new rays and improving overall throughput.

% \begin{figure}[htbp]
%     \centering
%     % \setlength{\abovecaptionskip}{0pt}
%     \includegraphics[width=\linewidth]{figs/wknd_active_warps.pdf}
%     % or figure* + \pagewidth for bigger images
%     \caption{Average number of warps per SM performing traversal for the WKND shadow scene. }
%     \label{fig:wknd_active_warps}
% \end{figure}

Among all scenes, WKND and SHIP show the least speedup. Table~\ref{tab:lumibench} presents the BVH tree depth and size for each scene. These two are the smallest scenes in the LumiBench suite. For WKND, the BVH tree size is only 0.2 MB, small enough to fit almost entirely in the L1 cache. The RT unit L1 miss rate confirms this: WKND SH achieves a miss rate of only \textbf{5.6\%}, and SHIP AO achieves \textbf{5.7\%}.
With L1 cache latency of only tens of cycles, these workloads become compute-bound rather than memory-bound. 
% However, these scenes are not representative of real-world workloads due to their small size. Even if such workloads existed in practice, Amdahl's Law dictates that accelerating traversal provides limited benefit when shaders dominate execution time.

\subsection{Memory}

% \begin{figure}[h]
% \centering
% % \setlength{\abovecaptionskip}{0pt}
% \includegraphics[width=\linewidth]{figs/l1_misses.pdf}
% \caption{{Normalized L1 and L2 cache misses. Only workloads with changes exceeding 10\% are shown. The geometric mean includes all workloads.}}
% \label{fig:l1_misses}
% \end{figure}

\begin{figure}[h]
    \centering
    \includegraphics[width=\linewidth]{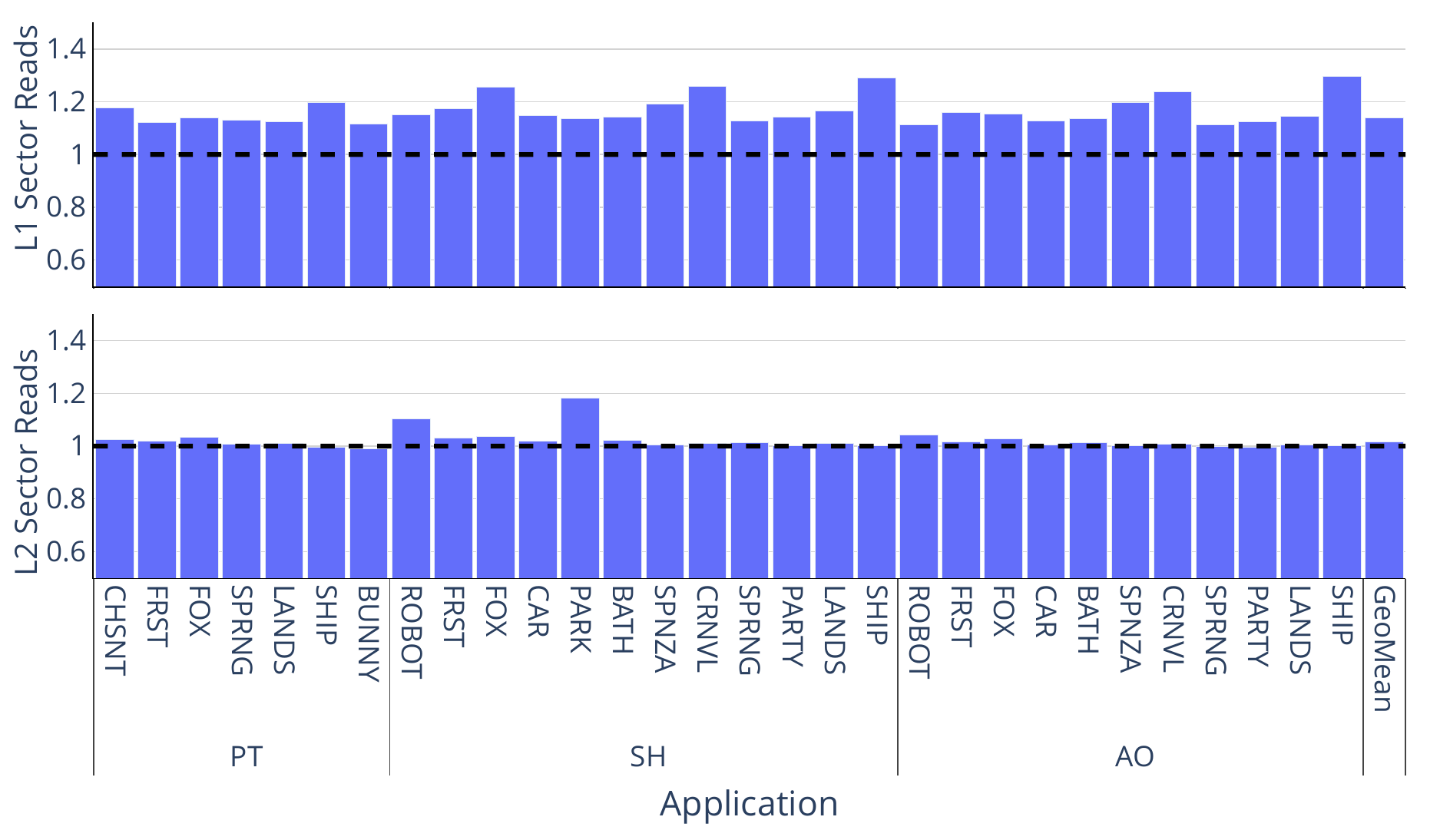}
    % or figure* + \pagewidth for bigger images
    \caption{L1  and L2 sector loads normalized to baseline. Only workloads with changes exceeding 10\% L1 accesses are shown. Geomean is across all workloads.}
    \label{fig:l1_l2_tot_sector}
\end{figure}

% However, enabling the simultaneous traversal of rays from different tiles alters the baseline memory traffic pattern. Because these bundled rays often traverse scattered paths through the BVH, they frequently lack spatial locality. 
{
During traversal, rays are grouped according to their placement in the ray buffer. By implicitly reordering rays within the RT unit, our design bundles rays from different warps to traverse concurrently. In the baseline architecture, execution divergence strictly limits the number of active rays, which reduces memory-level parallelism (MLP) and leaves performance strongly latency-bound. By sustaining more rays in flight, our design increases MLP and allows the GPU to hide memory latency more effectively, at the cost of added pressure on the memory subsystem. This pressure arises because implicit reordering can turn originally coalesceable warps more divergent, which generates redundant memory requests.
To quantify this effect, we analyze L1 and L2 sector load requests. Fig.~\ref{fig:l1_l2_tot_sector} reports L1 and L2 sector accesses normalized to the baseline, including only workloads with $>$10\% L1 variance. Across all workloads, we observe 14\% more L1 accesses on average. Because the proposed mechanism is constrained to the hardware traversal stage, this increase is the direct overhead of the implicit reordering introduced by decomposing the warp-granular buffers.
Although the reordering raises L1 traffic, L2 traffic stays nearly flat because the MSHRs absorb the redundant requests before they reach the L2. Two factors explain this. First, BVH node sizes exceed a single sector (Section~\ref{coalescing}), leaving no sub-sector coalescing for the reordering to disrupt, in contrast to the coalescing loss observed in general-purpose kernels \cite{fung_dynamic_2007}. Second, the redundant requests introduced by reordering target the same sectors, and the MSHRs coalesce these in-flight duplicates, leaving only one request per sector to propagate to the L2. As shown at the bottom of the figure, L2 sector accesses increase by only 1.6\% on average. 
}

\begin{figure}[h]
    \centering
    \setlength{\abovecaptionskip}{0pt}
    \includegraphics[width=\linewidth]{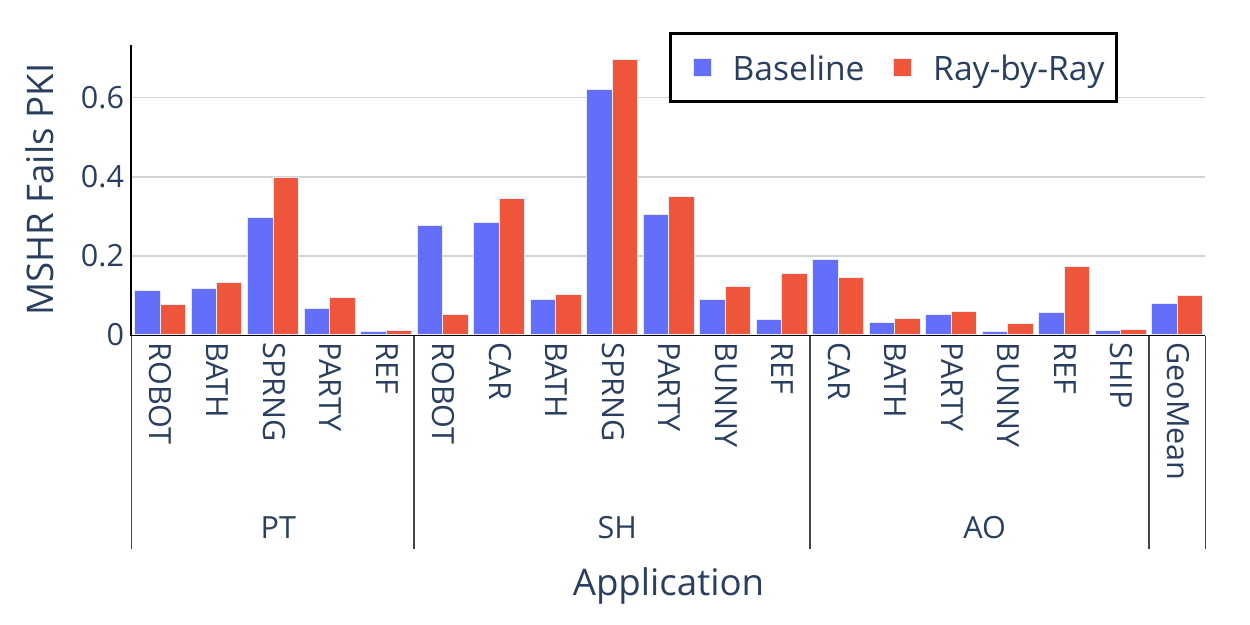}
    % or figure* + \pagewidth for bigger images
    \caption{{L1 MSHR merge fails per kilo instructions. Only workloads with changes exceeding 2\% are shown. Geomean is across all workloads.}}
    \label{fig:l1_mshr_merge_fail}
\end{figure}

{
While we rely on the MSHR merging capacity to absorb the extra L1 sector loads, we must confirm how the MSHR behaves and verify that we do not exhaust its capacity. The L1 data cache may stall for one of two reasons: no free MSHR entry is available, or an allocated entry cannot merge an additional request. Because the pointer-chasing nature of traversal bounds the total number of in-flight requests, we observe no MSHR allocation failure across any workload, in both the baseline and the proposed design.
Fig.~\ref{fig:l1_mshr_merge_fail} plots MSHR merge fails per kilo warp instructions. For space, we plot only applications whose merge fails change by more than 2\% relative to the baseline, while the geomean is computed across all workloads. At an average of only 0.09 fails per kilo warp instruction, the baseline L1 cache rarely stalls on the MSHR to begin with. Under the proposed design, the average stays at a negligible 0.10 fails per kilo warp instruction. 
}

{
These results confirm the picture from Section~\ref{coalescing}: the reordering raises L1 sector accesses, the MSHRs absorb the redundant inter-sector requests without exhausting capacity, and L2 traffic remains nearly unchanged.
}
\begin{figure}[h]
    \centering
    \setlength{\abovecaptionskip}{0pt}
    \includegraphics[width=\linewidth]{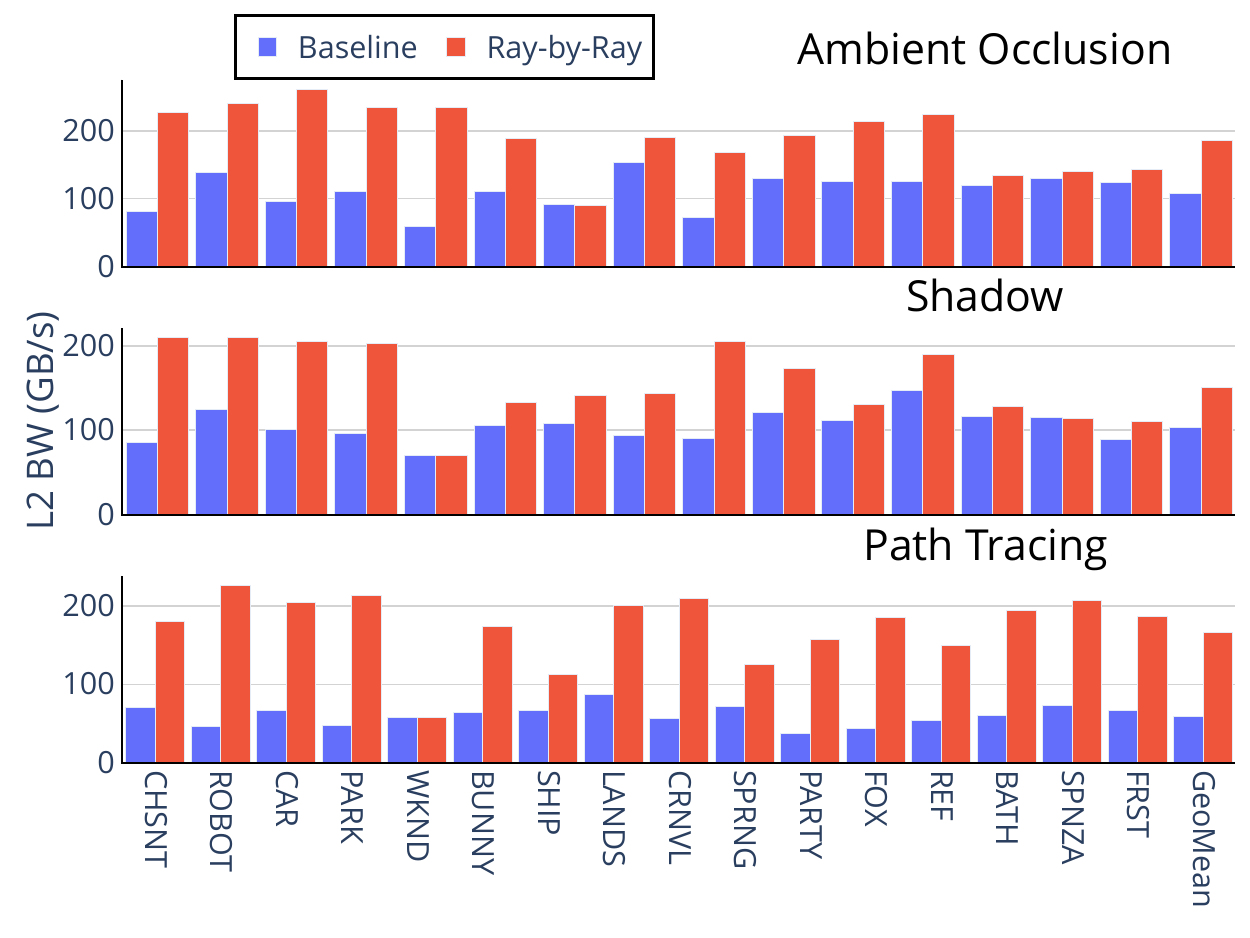}
    % or figure* + \pagewidth for bigger images
    \caption{{L2 bandwidth consumption.  Ray-by-Ray consumes 1.94$\times$ more L2 bandwidth than the baseline, reflecting the higher MLP from keeping more rays in flight.}}
    \label{fig:l2_bw}
\end{figure}

{
Achieved memory bandwidth is a useful proxy for overall throughput on a latency-hiding GPU. Fig.~\ref{fig:l2_bw} reports the L2 bandwidth of each design. Ray-by-Ray sustains 1.94$\times$ the baseline L2 bandwidth, reflecting the additional rays it keeps traversing in parallel. This bandwidth comes from useful work: Ray-by-Ray never spills to global memory and barely increases L1 misses.
Where L2 bandwidth is the binding constraint, this added demand can cap the achievable speedup. This is not a failure of the mechanism but a sign that the bottleneck has shifted to the L2. Reducing L2 traffic or expanding L2 bandwidth is orthogonal and complementary to our design.
}

% {
% We note one honest caveat. On a system where L2 bandwidth is the binding constraint, as can be the case on a larger desktop GPU, the achievable speedup may be capped by the available L2 bandwidth rather than by the RT unit. This does not mean the mechanism fails to work. On the contrary, it means we have exhausted the hardware resources required to saturate the RT unit, which is exactly the goal; the bottleneck simply shifts to the L2 and warrants separate attention. Such L2 bottlenecks are orthogonal to our design: any technique that reduces L2 traffic or expands L2 bandwidth is complementary to Ray-by-Ray and can be applied alongside it. Our design also does not create this pressure itself. The higher L2 bandwidth reflects more concurrent rays performing useful traversal, not inefficiency introduced by Ray-by-Ray: we never spill to global memory, and we observe virtually no increase in L2 misses (only 10.1\% additional L1 misses), so the design adds no extra L2 traffic beyond servicing the additional in-flight rays.
% }

\subsection{Energy} \label{energy}
\begin{figure}[h]
    \centering
    % \vspace{-5pt}
    % \setlength{\abovecaptionskip}{0pt}
    \includegraphics[width=\linewidth]{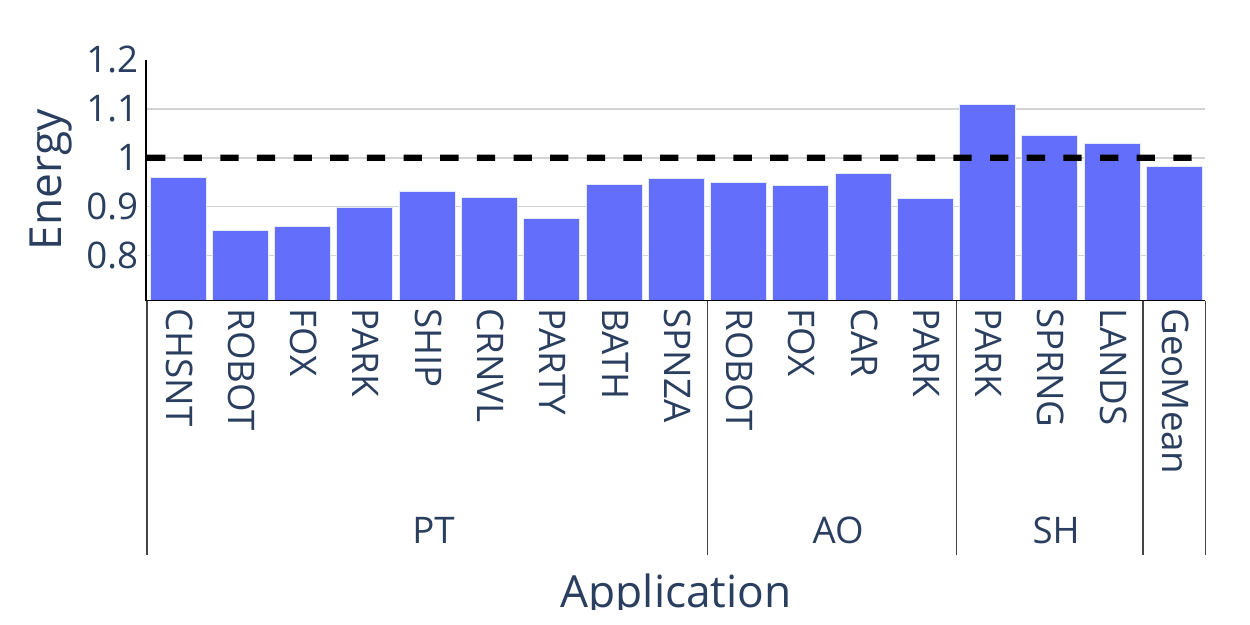}
    % or figure* + \pagewidth for bigger images
    \caption{Normalized energy consumption. Only workloads with changes exceeding 3\% are shown. The geometric mean includes all workloads.}
    \label{fig:energy}
\end{figure}

{We evaluated GPU energy consumption of the design with GPUWattch \cite{leng_gpuwattch_enabling_2013}.} Fig.~\ref{fig:energy} shows our design achieves an average energy consumption of 0.98$\times$ relative to the baseline. Because the design does not reduce the amount of work performed or the number of memory loads issued, large energy savings are not expected. However, the design completes traversal substantially faster, and the reduction in execution time naturally reduces energy consumption.

The largest reductions occur in PT workloads, which exhibit the highest levels of divergence and therefore benefit most from improved RT unit utilization. A few shadow workloads show increased energy consumption. 
As shown in Fig.~\ref{fig:l1_l2_tot_sector}, PARK~SH experiences the largest increase in off-chip accesses relative to the baseline, which is the primary cause of its higher energy usage. Even so, PARK~SH still achieves a 1.78$\times$ speedup, demonstrating that the additional energy cost is outweighed by the performance gain.

\subsection{Area Overhead}
As detailed in Section~\ref{design}, we decompose the monolithic warp-level buffers into individual ray buffer entries to enable fine-grained resource allocation. We evaluate hardware area overhead by synthesizing the RTL implementation using the Cadence Genus Synthesis tool (ver. 21.11) with a 15nm process design kit \cite{martins_open_cell_2015} at a 1GHz target clock frequency. The functional correctness of the design was verified through RTL-level simulations to ensure parity with our architectural model.
To ensure a fair area comparison, we configured the baseline RTL to allocate one thread per cycle, matching the serialized allocation logic of our proposed design. Importantly, this serialization is applied only for area synthesis; the baseline retains its original multi-lane allocation width in our performance evaluation.
Consistent with the baseline design \cite{saed_vulkansim_2022, tozlu_cooprt_2025, chou_treelet_prefetching_2023}, we allocate 290 bits of metadata per ray entry. Based on our characterization in Section~\ref{stack_management}, we assume the traversal stack is stored in shared memory. This differs from prior works \cite{saed_vulkansim_2022, liu_intersection_prediction_2021, chou_treelet_prefetching_2023} that assume large on-chip stack storage, which can account for approximately 70\% of the total bytes stored within the unit; under such a baseline, the relative overhead of our design would be further reduced.

\begin{figure}[h]
    \centering
    \includegraphics[width=\linewidth]{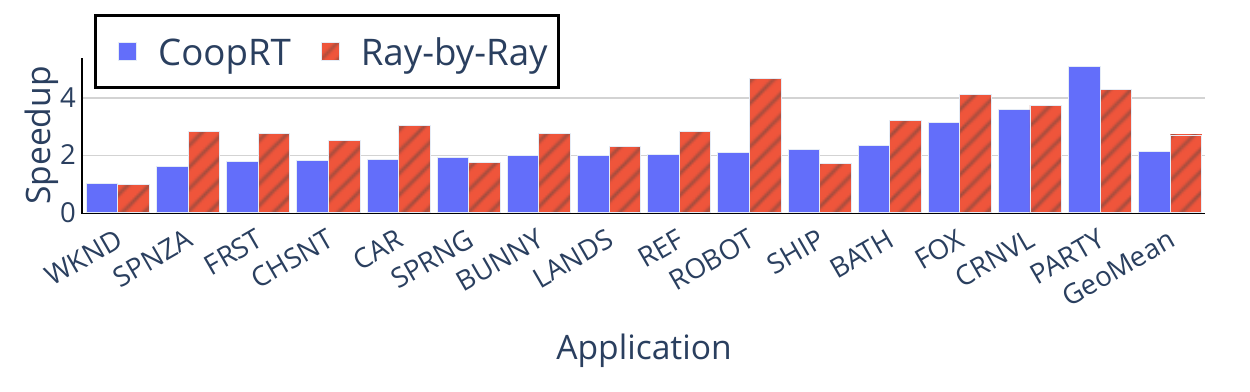}
    % or figure* + \pagewidth for bigger images
    \caption{Comparing proposed design with CoopRT for PT workloads. }
    % \vspace{-5pt}
    \label{fig:cooprt}
\end{figure}

\begin{figure*}[htbp]
    \centering
    \includegraphics[width=\linewidth]{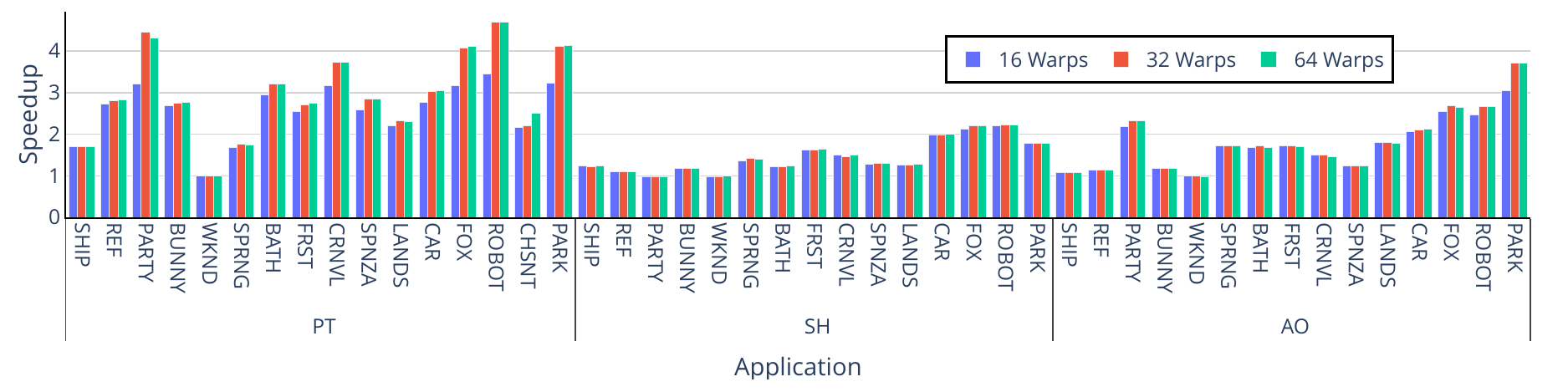}
    % or figure* + \pagewidth for bigger images
    \caption{Normalized speedup when limiting the number of warps in RT unit.}
    \label{fig:warp_count_speedup}
\end{figure*}

To optimize storage, we limit the warp table to 32 entries. Our analysis in Fig.~\ref{fig:warp_count_speedup} demonstrates that increasing warp capacity from 32 to 64 yields negligible speedup, as the pipelined nature of traversal prevents all warps from performing active traversal simultaneously. This configuration requires 32 completion counters (5 bits each) and a 5-bit Warp ID field for each of the 128 ray buffer entries. Including a 128-bit occupancy mask, the total additional bookkeeping state is 928 bits ($128 + 5 \times 32 + 5 \times 128$). Our RT core model incorporates eight traversal math pipelines, following the throughput analysis and design specifications established in RayFlex \cite{shen_rayflex_opensource_2025}. Synthesis results show the baseline RT core occupies 677,066.8 $\mu m^2$, while our improved RT core occupies 678,368.4 $\mu m^2$. Thus, the proposed fine-grained management introduces a minimal 0.19\% total area overhead.
{
% To further characterize the microarchitectural implementation, we performed SAIF-annotated dynamic power analysis. We selected representative operating points from Lumibench statistics to cover a range of conditions, such as 30\% RT efficiency, 41 nodes traversed (the average across Lumibench), and node load latency assuming a 50\% L1 hit rate. Averaged across all of these points, the baseline and Ray-by-Ray designs consume 9.17 mW and 9.23 mW, respectively, confirming negligible power overhead of less than 0.07\%.
GPUWattch (Section~\ref{energy}) models system-level GPU power but does not instrument the internal logic of the RT unit. The normalized energy in Fig.~\ref{fig:energy}, therefore, reflects changes in overall GPU system-wise activity, but it does not capture the incremental power of the added control structures that Ray-by-Ray places inside the RT unit.
Because the RTL synthesis above already produces a gate-level netlist, we reuse that artifact to measure RT-unit power directly, a quantity GPUWattch does not capture. We sweep a range of operating points drawn from LumiBench statistics to span realistic conditions, varying RT efficiency, traversal node count, and node load latency, and perform SAIF-annotated dynamic power analysis at each. Across all tested points, the baseline and Ray-by-Ray designs consume 9.17~mW and 9.23~mW on average, respectively, a negligible 0.7\% overhead.}

\subsection{Comparison To Related Work} \label{cooprt}
CoopRT~\cite{tozlu_cooprt_2025} improves traversal efficiency by using inactive threads as helpers to parallelize BVH traversal for the active ray. The technique is evaluated on PT workloads. Its approach focuses on intra-warp optimization: it splits traversal work within a warp so that helper lanes can explore different parts of the tree in parallel, accelerating a single ray’s progress. In contrast, our work targets inter-warp parallelism by managing RT-core resources more efficiently, allowing more warps to traverse concurrently and improving overall throughput.

Fig.~\ref{fig:cooprt} compares our proposed design with CoopRT. The CoopRT results are taken directly from the authors’ artifact evaluation and therefore reflect a different experimental configuration (RTX 2060).
Ray-by-Ray outperforms CoopRT on most workloads. In a standard BVH traversal, many branches are pruned early when their intersection distance exceeds the ray’s current $t_{\min}$. Since CoopRT explores multiple branches in parallel before $t_{\min}$ is updated, some lanes may traverse subtrees that would have been pruned in a sequential traversal.
In the few cases where CoopRT outperforms our design, the workloads exhibit extreme divergence with very few active threads. Although our mechanism enables all pending warps to perform traversal, the available ray concurrency remains insufficient to fully saturate the RT unit.
% CoopRT leverages intra-warp parallelism and is therefore more effective when divergence is extremely high and ray-level parallelism is limited. 
CoopRT leverages intra-warp parallelism and is therefore more effective in these cases.
Nevertheless, our design achieves a higher geometric mean speedup across all workloads.

% \subsection{initiation intervals} 

% In Section~\ref{design} we explained the initiation interval of the proposed RT unit is longer because each ray needs to be checked annd allocated a ray buffer. To quantify the that, we apply the long initiation interval to both the baseline design and the proposed design. 

\section{Related Work}

\textbf{{Ray Reordering: }}
{A natural way to mitigate the observed low RT efficiency is to reorder rays into more coherent groups. This can be done through software ray sorting \cite{meister_ray_reordering_2025,moon_cacheoblivious_ray_2010,t.reis_coherent_rayspace_2017,szecsi_hierarchical_ray_,costa_accelerating_occlusion_2015}, or, most prominently, through NVIDIA's Shader Execution Reorder (SER) \cite{nvidia_shader_execution_}, which allows developers to define explicit reordering points within a shader to regroup threads into more coherent warps using hints such as hit material or ray activity.
While SER is most often associated with material-shading coherence, it is not limited to it: in path tracing workloads, developers can use the ray active flag as the sorting key to regroup surviving rays between bounces \cite{nvidia_path_tracing_2025}.}

However, these approaches do not resolve the fundamental hardware inefficiencies described in this work, for two primary reasons. First, reordering incurs significant performance overhead due to the state that must be spilled between stages. As demonstrated in the path tracing optimizations for \textit{Indiana Jones and the Great Circle}, the effectiveness of SER is heavily constrained by the volume of live-state bytes that must be spilled to memory \cite{nvidia_path_tracing_2025}. {Regrouping rays requires this live state to be written out and reloaded after reordering, which increases L2 cache traffic and introduces additional memory latency \cite{nvidia_improve_shader_2022}. On contemporary GPUs with hardware-accelerated traversal, this overhead is large enough that reordering often fails to pay off \cite{meister_ray_reordering_2025,kim_cost_divergence_2023}. }
Second, even where reordering reduces inter-path divergence, because it is a software-driven mechanism that operates at the shader level, it cannot address the ``long-tail'' effect occurring within the monolithic traversal instruction \cite{muthler_techniques_traversing_2022}. Since the RT core manages buffers and traversal stacks at warp granularity, these on-chip resources remain occupied until the entire warp commits traversal, even if only a single thread remains active. Furthermore, as a strictly post-traversal mechanism, reordering such as SER cannot mitigate the \textit{inter-ray} (intra-warp) divergence that arises while rays are actively traversing the scene. Consequently, these techniques cannot provide the fine-grained, intra-instruction resource reclamation necessary to overcome the underlying warp-level resource under-utilization.

\textbf{SIMT Efficiency: }
A large body of prior work investigates execution divergence and techniques to improve SIMT efficiency on GPUs. Dynamic Warp Formation~\cite{fung_dynamic_2007} and Thread Block Compaction~\cite{fung_thread_block_2011a} regroup threads at divergent branch points to increase warp coherence. 
These approaches require explicit tracking of divergence and convergence points, complicating hardware design. In contrast, ray traversal naturally provides well-defined divergence boundaries: threads enter the monolithic traversal instruction together and reconverge only when it completes.
Traditional vector-SIMD pipelines also require lane data to remain ordered, making lane migration expensive. This constraint does not apply to RT traversal. Ray state is stored within dedicated RT-unit structures rather than the vector register file, and the datapath is much narrower, allowing threads from different warps to be packed more efficiently.
Prior techniques may also introduce additional memory-system overheads, as reorganizing threads can produce more divergent accesses and reduce coalescing efficiency. As shown in our analysis, coalescing offers little benefit during BVH traversal, and our design builds on this insight.
Aila et al.~\cite{aila_understanding_2009} implemented a GPU ray tracer using persistent threads, where terminated rays are periodically replaced with new ones to mitigate SIMT underutilization. However, this approach is not applicable to a dedicated RT unit, as traversal executes as a single monolithic instruction and thread contexts cannot be modified until the instruction completes.

\textbf{GPU Ray Tracing: }
Recent works~\cite{kopta_efficient_mimd_2010, spjut_trax_multicore_2009} propose MIMD-threaded processors for performing traversal. Other architectures, such as SGRT~\cite{lee_sgrt_mobile_2013} and RayCore~\cite{nah_raycore_raytracing_2014}, target ray tracing on mobile devices. 
These works focus on the traversal mechanism itself rather than the higher-level inefficiencies addressed in this paper.
% The key distinction lies in the mismatch between a monolithic, warp-level traversal instruction and a thread-level MIMD execution model.
CoopRT~\cite{tozlu_cooprt_2025} attempts to improve traversal efficiency by using inactive lanes to assist with BVH traversal for the active ray. However, as explained in Section~\ref{cooprt}, this technique may perform unnecessary work, which can limit the achievable speedup.
% While the approach is functionally correct, it can occasionally perform unnecessary traversal work. In a standard BVH traversal, many branches can be pruned early when their intersection distance exceeds the ray’s current $t_{\min}$. Because CoopRT explores multiple branches in parallel before $t_{\min}$ is updated, some lanes may traverse subtrees that would have been pruned in a sequential traversal.

Many prior works aim to improve the microarchitecture of ray tracing accelerators, and most of these techniques are orthogonal to our design and can be integrated alongside it. Liu et al.~\cite{liu_intersection_prediction_2021} proposed caching intersection results and using ray-property hashing to predict future intersections, thereby avoiding unnecessary traversal. Their design reconstructs warps during traversal to mitigate misprediction overhead but still manages ray buffers at warp granularity and does not increase the number of warps that can traverse concurrently.
Jung et al.~\cite{jung_hierarchical_2025} use shared memory as auxiliary storage for traversal stack overflows. Although we assume that the short stack resides entirely in shared memory, the shared-memory management techniques proposed in this work can be applied to their design as well.
Chou et al.~\cite{chou_treelet_prefetching_2023} propose prefetching BVH treelets to reduce memory latency during traversal.
A large body of work~\cite{pharr_rendering_complex_1997,garanzha_fast_ray_2010,moon_cacheoblivious_ray_2010,ylitie_efficient_incoherent_2017,navratil_dynamic_ray_2007} explores ray sorting to improve memory locality and reduce divergence during GPU-based traversal. 
% These works focus on accelerating the traversal algorithm, whereas our work targets on-chip runtime resource management to improve traversal throughput and enable more warps to traverse in parallel.

\section{Conclusion}
We presented Ray-by-Ray, a new mechanism for efficient on-chip resource management in hardware-accelerated GPU ray tracing units. Our analysis reveals a mismatch between the GPU’s SIMT model and the per-thread execution model of RT units, leading to inefficient warp-level resource allocation. Inactive lanes continue to occupy buffer entries, reducing parallelism and limiting traversal throughput. Ray-by-Ray introduces a lightweight, fine-grained per-ray allocation scheme that tracks only active lanes, enabling more rays to traverse concurrently and better hiding memory latency. We also show that contemporary GPU memory coalescing provides limited benefit for ray tracing, allowing relaxed warp-level coalescing constraints.

We implemented the Ray-by-Ray architecture in RTL and evaluated its performance across diverse workloads and ray tracing techniques using Vulkan-Sim and LumiBench. Our design achieves an average speedup of 1.91$\times$ while incurring a minimal total area overhead of 0.19\%.

\section*{Acknowledgment}
The authors would like to thank Wilson Wai Lun Fung for his mentorship, insightful discussions, and brainstorming during the early stages of this work, which inspired the core ideas of this paper. This work was supported, in part, by NSF CCF \#1943379 and NSF CCF \#2450447. We also thank the anonymous reviewers for their constructive feedback.
% ChatGPT and Claude were used throughout the preparation of this manuscript for writing assistance, title suggestion, and grammar checking. Cursor, Claude Code, and Copilot were used to support code implementation. All AI-assisted content was fully reviewed and verified by the authors, and all scientific ideas, contributions, and conclusions are original to the authors.

%%%%%%% -- PAPER CONTENT ENDS -- %%%%%%%%

%%%%%%%%% -- BIB STYLE AND FILE -- %%%%%%%%
\bibliographystyle{IEEEtranS}
\bibliography{references}
%%%%%%%%%%%%%%%%%%%%%%%%%%%%%%%%%%%%

\appendix[Artifact Appendix]

\subsection{Abstract}

This artifact reproduces Fig.~8 of the paper --- the per-workload
normalized speedup of the Ray-by-Ray RT unit over the baseline warp-granular
RT unit, across 46 ray-tracing workloads
(16 scenes from LumiBench $\times$ SH/AO/PT shader modes, plus \texttt{CHSNT\_PT}).
It is packaged as two self-contained Docker images (design and baseline)
built on Vulkan-Sim; the baseline sweep results are pre-computed and shipped
inside the tarball, so the reviewer only needs to run the design sweep and
render the figure.

\subsection{Artifact check-list (meta-information)}

{\small
\begin{itemize}
  \item {\bf Program:} Vulkan-Sim v2.0 (cycle-level GPGPU-Sim + Mesa/lavapipe)
        running the RayTracingInVulkan benchmark app on the LumiBench scenes.
  \item {\bf Compilation:} Pre-built inside the shipped Docker images
        (Ubuntu 24.04, gcc-11, CUDA~11.7 \texttt{ptxas}, LLVM-11).
  \item {\bf Binary:} Simulator, Vulkan driver, benchmark app, scenes, and
        toolchain are all baked into the images.
  \item {\bf Data set:} LumiBench (16 scenes, included).
  \item {\bf Run-time environment:} Linux host with Docker~$\geq$~20.10.
  \item {\bf Hardware:} Any x86-64 host with $\geq$~128~GB of RAM. No GPU
        required. More RAM and more CPU cores directly reduce wall time.
  \item {\bf Metrics:} \texttt{gpu\_tot\_sim\_cycle} per workload; the plot
        reports per-workload speedup (baseline$/$design) and their geomean.
  \item {\bf Output:} \texttt{speedup.png} / \texttt{speedup.pdf}
        (Fig.~8), plus per-run Vulkan-Sim logs.
  \item {\bf Experiments:} One design-sweep run reproduces Fig.~8.
  \item {\bf How much disk space required (approximately)?:}
        ~15~GB (images + shipped baseline results + design sweep outputs).
  \item {\bf How much time is needed to prepare workflow (approximately)?:}
        $<$10~min (\texttt{docker load} + \texttt{tar -xf}).
  \item {\bf How much time is needed to complete experiments (approximately)?:}
        ~1~week serial; $\sim$1~day when run in parallel on a well-provisioned
        host (higher RAM/cores $\Rightarrow$ more concurrent workloads).
  \item {\bf Publicly available?:} Yes, on Zenodo.
  \item {\bf Code licenses (if publicly available)?:} BSD (inherited from
        Vulkan-Sim / GPGPU-Sim).
  \item {\bf Archived (provide DOI)?:} 10.5281/zenodo.21542536.
\end{itemize}
}

\subsection{Description}

\subsubsection{How to access}

A single tarball on Zenodo (DOI above) containing: (i) the artifact source
tree (mirror of the working directory, minus intermediate outputs), (ii) the
pre-built Docker image tarball \texttt{vulkan-sim-artifact.tar}
(\texttt{rbr-design} and \texttt{rbr-baseline}), (iii) the pre-computed
baseline sweep results under \texttt{results/baseline/}, and (iv) the driver
scripts under \texttt{docker/scripts/}.

\subsubsection{Hardware dependencies}

x86-64 Linux host with Docker and $\geq$~128~GB RAM. No GPU is required
(rendering is CPU-emulated through Mesa/lavapipe; GPGPU-Sim is a CPU
simulator). One workload, \texttt{CHSNT\_PT}, has a peak resident set
of $\sim$120~GB and will be skipped automatically on hosts with less
headroom. Wall time scales with available RAM and CPU cores because the
in-container scheduler bin-packs workloads against a memory budget.

\subsubsection{Software dependencies}

Docker only. All simulator, driver, toolchain, Python plotting, and benchmark
dependencies are contained inside the images.

\subsubsection{Data sets}

LumiBench (16 scenes) is shipped inside the images. Rendering resolution
is 256$\times$256, matching the paper.

\subsection{Installation}

From the extracted tarball root:

{\footnotesize
\begin{verbatim}
docker load -i vulkan-sim-artifact.tar
\end{verbatim}
}

This registers \texttt{rbr-design} and \texttt{rbr-baseline}. The
pre-computed baseline results are already in \texttt{./results/baseline/}.

\subsection{Experiment workflow}

Run the design sweep and render Fig.~8:

{\footnotesize
\begin{verbatim}
# defaults: use all cores, all RAM
./docker/scripts/make_speedup.sh --only design

# cap concurrency and per-sweep memory
# budget explicitly
JOBS=8 ./docker/scripts/make_speedup.sh \
    --only design --max-mem 200
\end{verbatim}
}

This launches a single \texttt{rbr-design-sweep} container that runs all 46
workloads in parallel inside the container (GNU \texttt{parallel} + a memory
reservation scheduler), then invokes the in-image plotting script against
\texttt{results/design/} (freshly generated) and \texttt{results/baseline/}
(shipped). \texttt{JOBS} caps the concurrent-workload count (default:
container CPU count). \texttt{--max-mem GB} caps the total memory budget the
scheduler bin-packs against; any workload whose reservation exceeds the
(auto-clamped) budget is skipped rather than launched.

\paragraph{Other entry points.}
See \texttt{docker/README.md} for the full script surface. In brief:
\begin{itemize}
  \item \texttt{run\_benchmark.sh} --- run a single workload;
        \texttt{-{}-list} enumerates them.
  \item \texttt{run\_sweep.sh} --- the in-container parallel driver
        invoked by \texttt{make\_speedup.sh}; can also be run
        standalone to inspect the sweep container directly.
  \item \texttt{run\_all\_parallel.sh} --- host-side alternative that
        launches one container per benchmark instead of one per variant.
\end{itemize}

\subsection{Evaluation and expected results}

The script produces \texttt{./speedup.png} and \texttt{./speedup.pdf}. The
figure should reproduce Fig.~8: per-application, per-shader-mode
(SH/AO/PT) speedup bars, with the overall geometric mean and the relative
ranking across workloads matching the paper. Small numerical variation in
individual bars is expected and does not affect the qualitative result.

\subsection{Experiment customization}

To sweep at a lower resolution (much faster, lower fidelity):

{\footnotesize
\begin{verbatim}
WIDTH=64 HEIGHT=64 \
    ./docker/scripts/make_speedup.sh \
    --only design
\end{verbatim}
}

To re-run the baseline sweep from scratch instead of using the shipped
results, drop the \texttt{--only design} flag.

\subsection{Notes}

\paragraph{Baseline vs. upstream Vulkan-Sim.}
The \texttt{rbr-baseline} image is essentially original Vulkan-Sim 2.0 (commit 
dc742fa). The
differences from public upstream are (a) additional performance counters
used elsewhere in the paper's characterization and (b) software
optimizations to reduce simulation wall time.

\paragraph{Benign teardown abort after each run.}
After a workload completes, the reviewer may see something like:
{\footnotesize
\begin{verbatim}
GPGPU-Sim: *** exit detected ***
RayTracer: .../util/u_surface.c:267:
  util_resource_copy_region:
  Assertion `src_bs == dst_bs' failed.
Aborted (core dumped)
\end{verbatim}
}
This is benign. The assertion fires in Mesa/lavapipe's Vulkan teardown
\emph{after} GPGPU-Sim has already printed its exit marker and flushed all
stats to the log; the simulation itself completed successfully. The
authoritative signal is the line
``\texttt{GPGPU-Sim: *** exit detected ***}'' plus a
\texttt{gpu\_tot\_sim\_cycle} value in the per-workload \texttt{.log}.

\end{document}